\documentclass[twocolumn]{aastex631}
\pdfoutput=1 
\usepackage{amsmath,amstext,amssymb,mathtools,graphicx,color}

\usepackage{graphicx}	
\usepackage{amsmath}	
\usepackage{amssymb}	
\usepackage{xcolor}
\usepackage{lipsum}
\usepackage[normalem]{ulem}

\submitjournal{ApJ}

\shorttitle{BLSN}
\shortauthors{Fryer et al.}

\graphicspath{{./}{figures/}}

\begin{document}

\title{Explaining Non-Merger Gamma-Ray Bursts and Broad-Lined Supernovae with Close Binary Progenitors with Black Hole Central Engines}
d
\correspondingauthor{Chris Fryer}
\email{fryer@lanl.gov}

\author[0000-0003-2624-0056]{Christopher L. Fryer}
\affiliation{Center for Nonlinear Studies, Los Alamos National Laboratory, Los Alamos, NM 87545 USA}

\author[0000-0002-2942-3379]{Eric Burns}
\affiliation{Department of Physics \& Astronomy, Louisiana State University, Baton Rouge, LA 70803, USA}

\author[0000-0002-9017-3567]{Anna Y. Q. Ho}
\affiliation{Department of Astronomy, Cornell University, Ithaca, NY 14853, USA}

\author[0000-0001-8104-3536]{Alessandra Corsi}
\affiliation{William H. Miller III Department of Physics \& Astronomy, Johns Hopkins University, 3400 N. Charles Street
Baltimore, MD 21218}

\author[0000-0002-7851-9756]{Amy Y. Lien}
\affiliation{Department of Chemistry, Biochemistry, and Physics, University of Tampa, 401 W. Kennedy Boulevard, Tampa, FL 33606, USA}

\author[0000-0001-8472-1996]{Daniel A. Perley}
\affiliation{Astrophysics Research Institute, Liverpool John Moores University, IC2, Liverpool L3 5RF, UK}

\author{Jada L.~Vail}
\affiliation{Department of Astronomy, Cornell University, Ithaca, NY 14853, USA}

\author[0000-0002-5814-4061]{V.~Ashley Villar}
\affiliation{Center for Astrophysics \textbar{} Harvard \& Smithsonian, 60 Garden Street, Cambridge, MA 02138-1516, USA}
\affiliation{The NSF AI Institute for Artificial Intelligence and Fundamental Interactions}

\begin{abstract}
For over 25 years, the origin of long-duration gamma-ray bursts (lGRBs) has been linked to the collapse of rotating massive stars. However, we have yet to pinpoint the stellar progenitor powering these transients. Moreover, the dominant engine  powering the explosions remains open to debate.  Observations of both lGRBs, supernovae associated with these GRBs, such as broad-line (BL) stripped-envelope (type Ic) supernovae (hereafter, Ic-BL) supernovae (SNe) and perhaps superluminous SNe, fast blue optical transients, and fast x-ray transients, may provide clues to both engines and progenitors.  In this paper, we conduct a detailed study of the tight-binary formation scenario for lGRBs, comparing this scenario to other leading progenitor models.  Combining this progenitor scenario with different lGRB engines, we can compare to existing data and make predictions for future observational tests.  We find that the combination of the tight-binary progenitor scenario with the black hole accretion disk (BHAD) engine can explain lGRBs, low-luminosity GRBs, ultra-long GRBs, and Ic-BL. We discuss the various progenitor properties required for these different subclasses and note such systems would be future gravitational wave merger sources. We show that the current literature on other progenitor-engine scenarios cannot explain all of these transient classes with a single origin, motivating additional work. We find that the tight-binary progenitor with a magnetar engine is excluded by existing observations. The observations can be used to constrain the properties of stellar evolution, the nature of the GRB and the associated SN engines in lGRBs and Ic-BL.  We discuss the future observations needed to constrain our understanding of these rare, but powerful, explosions.
\end{abstract}


\section{Introduction} 
\label{sec:intro}

Gamma-ray bursts (GRBs) are the most luminous events in the universe.  Although they were discovered over 50 years ago~\citep{1973ApJ...182L..85K}, the true power of these explosions was not fully understood until distance measurements allowed quantification of their intrinsic brightness~\citep{1997Natur.387..878M,1997Natur.389..261F,1998ApJ...507L..25B}.  While the community quickly converged on a narrow set of possible long duration gamma-ray bursts (lGRBs) engine and progenitor models, their exact properties are still not fully understood.  With better studies of different progenitors and engines, we can both learn more from observations of these explosions and better understand their role in the menagerie of transient outbursts observed in the universe.

Proposed GRB engines fall into three major engine paradigms~\citep{2019EPJA...55..132F}:  a) black hole accretion disk engines (BHAD) powered either by energy in the disk or the rotating black hole~\citep{1993ApJ...405..273W,1999ApJ...518..356P}, b) magnetar engines powered by the rotational energy in a spinning neutron star~\citep[NS;][]{2000ApJ...537..810W,2001ApJ...552L..35Z}, or c) NS accretion disk (NSAD) powered by the energy in a disk around a NS~\citep{1985ApJ...290..721M}.  The progenitors of these engines are very similar.  Most progenitors are either the mergers of compact objects (e.g. NS mergers with black holes, NSs or white dwarfs; for the magnetar engine, this includes white dwarf/white dwarf mergers) or the collapse of the cores of massive stars from either massive stars or through the merger of a compact object with a massive star~\citep{1999ApJ...526..152F}.   

\cite{2007PASP..119.1211F} summarized a community consensus of the different progenitor scenarios of lGRBs, reviewing many of the relevant observations constraining the GRB progenitor scenario, including the rate, associated supernova (SN), metallicity, surrounding environment, host type, and distribution within the host (offsets).  The progenitors considered spanned many of the leading progenitors proposed at the time.  These included single stars, both with normal stellar mixing parameters~\citep{1993ApJ...405..273W,1999ApJ...526..152F} and with extended mixing, a.k.a. homogeneous stars~\citep{2005A&A...443..643Y}.  A number of binary progenitors were also considered (many from the previous extensive study by \cite{1999ApJ...526..152F}) including short period binaries allowing tidal spin-up~\citep{2007Ap&SS.311..177V} and a broad set of merger scenarios~\citep{2003fthp.conf...19I,2005ApJ...623..302F} including mergers with a single compact object~\citep{1998ApJ...502L...9F}.  The Binary-driven hypernova model~\citep{2012ApJ...758L...7R,2014ApJ...793L..36F} required tight binaries to drive the accretion-induced collapse, arguing that the supernova would be from a stripped star (type Ic).  A set of cluster formation scenarios were also studied~\citep{2005Ap&SS.300..247P}.  For the most part, this study reviewed the different scenarios.  None of the progenitors studied by \cite{2007PASP..119.1211F} fit the observations perfectly.  Because of this, and because there were concerns that the observational interpretations could be wrong, these progenitors persisted.

For example, most progenitor models foresee a GRB jet propagating through a wind circumstellar medium, but a large fraction of lGRB observations seem to be better fit by a constant density interstellar medium (ISM) profile~\citep{2000ApJ...536..195C,2005MNRAS.363.1409P}.  While expansion in the ISM would argue against massive star progenitors, the GRBs appeared to occur in star-forming galaxies and star-forming regions~\citep{2010MNRAS.405...57S}. Hence, it was assumed that approximations in the afterglow models may lead to a misinterpretation of the observations as favoring the ISM constant density profile.  As we produce more accurate afterglow models and build better progenitor models (including a better understanding of stellar mass-loss), afterglow  observations may provide crucial input to the progenitors (and stellar mass-loss itself).

The currently-favored engine and progenitor for lGRBs is the collapsar model~\citep{1993ApJ...405..273W} that invokes the collapse of the rotating core of a massive star. Note that throughout this paper we are focusing on lgrbs from collapsars, and neglecting the contribution of lgrbs which arise from mergers \citep{rastinejad2022kilonova,levan2024heavy}.  The classic collapsar model relies on the BHAD engine and a number of potential progenitors to produce the required rotation speeds~\cite[for reviews, see][]{1999ApJ...526..152F,2007PASP..119.1211F}.  Because GRBs are rare, it is difficult to identify the exact progenitor or progenitors of this engine for long duration bursts.  Less than 1 in 1000 of all stellar collapses produce GRBs~\citep{2014ApJ...783...24L,2020ApJ...904...35P}.  This rate is uncertain, depending on both estimates of the total observed rate and the beaming angle~\citep{2015PhR...561....1K}. Roughly 1 in 10 stellar collapses make black holes~\citep{2001ApJ...554..548F}, so this means that less than 1\% of all black hole forming systems are needed to explain the bulk of all lGRBs. With such a small fraction of systems producing these explosions, a wide range of, at times exotic, progenitor scenarios are possible.  If we include other engine scenarios (e.g. magnetar or NSAD), identifying the exact progenitor proves even more difficult.

One of the key observations that provide insight into the progenitors producing lGRBs is the nature of the SNe often associated with lGRBs~\citep{Woosley_2006,Cano_2017}.  Initially called hypernovae~\citep{1998Natur.395..672I}, the broad-line (BL) features from the high-velocity Doppler broadening of these explosions~\citep{2002ApJ...572L..61M} led to the adoption of the BL SN nomenclature.  Hypernovae or BL SNe refer to any SNe with high-velocity (15,000--20,000\,km/s) features at the peak of the optical light curve, regardless of whether they are associated with GRBs.  Thusfar, the observed BL SNe are all type Ic i.e., stripped-envelope with little or no evidence of H or He lines \citep{2016ApJ...832..108M}.  
Hence, progenitor models for these Ic-BL SNe require the SN explosion to either prevent any helium emission (e.g. by completely ionizing the helium) or shed the helium before the explosion.

There is growing evidence that Ic-BL SNe, whether or not associated with GRBs, are produced by the same GRB engine/progenitor scenario, e.g. \cite{2017MNRAS.472..616S,Barnes_2018,2020ApJ...892..153M}, (more on this is section~\ref{sec:SN}).  If we assume that all Ic-BL SNe are produced by the same engine (e.g. BHAD, magnetar, etc.) mechanism, we can use the entire population of Ic-BL SNe to probe the nature of the progenitor and engine.  Viewing angle effects could explain the diversity of GRB strengths and their relation to Ic-BL SNe~\citep{2018ApJ...860...38B}.  In this paper, we instead assume that differences in the duration and power of the engine (caused by differences in the progenitor) explain not only the different types of long-duration bursts but also Ic-BL supernovae with and without GRBs.

In this paper, we focus on one of the leading progenitor scenarios of lGRBs, namely, tidally spun-up binaries.  One of the difficulties in this study is the range of results from stellar codes, and we consider a range of stellar models in our study of angular momenta and compact remnant spins (Section~\ref{sec:spins}).  We compare the results from these tidally-locked binaries (we include both He- and CO-star binaries) with those of single stars.  With these angular momentum results, we then study the predictions of the magnetar model (Section~\ref{sec:magnetar}) and the BHAD model using a number of mechanisms producing the jet (Section~\ref{sec:collapsar}).  This analysis refines the predictions of our tidally-locked binary scenario for the properties of GRBs and SNe (Section~\ref{sec:theoryobs}).  These predictions are then compared to SN (particularly Ic-BL) observations in Section~\ref{sec:GRB} and GRB properties in Section~\ref{sec:SN}.  In each of these two sections, we both compare predictions to current observations and argue for future observations that will further constrain our engine and progenitor models.  We conclude with a review of our results and a comparison to other potential progenitors.

\section{Angular Momentum and Compact Remnant spins}
\label{sec:spins}

Whether the lGRB engine is a magnetar, a NSAD, or a BHAD, the collapsing star must have considerable angular momentum to produce the powerful jets observed in GRBs.  This is difficult to achieve in single star models.  The problem arises from the fact that when massive stars expand off the main sequence, angular momentum conservation causes their spin rate to decrease.  If the stellar core is coupled to the envelope such that its spin is set to the envelope spin, the collapsed core will not have enough angular momentum to drive a GRB engine.  A number of scenarios have been proposed to either retain the high spin rate or spin up the star~\citep{1999ApJ...526..152F}.  A broad range of stellar and compact remnant observations (e.g. pulsars, X-ray binaries, and merging black hole systems) have been leveraged to provide clues into the nature of the spins in stellar cores~\citep{2020A&A...636A.104B}.  These observations suggest a wide range of core spin periods.

\begin{figure}
    \includegraphics[width=0.5\textwidth]{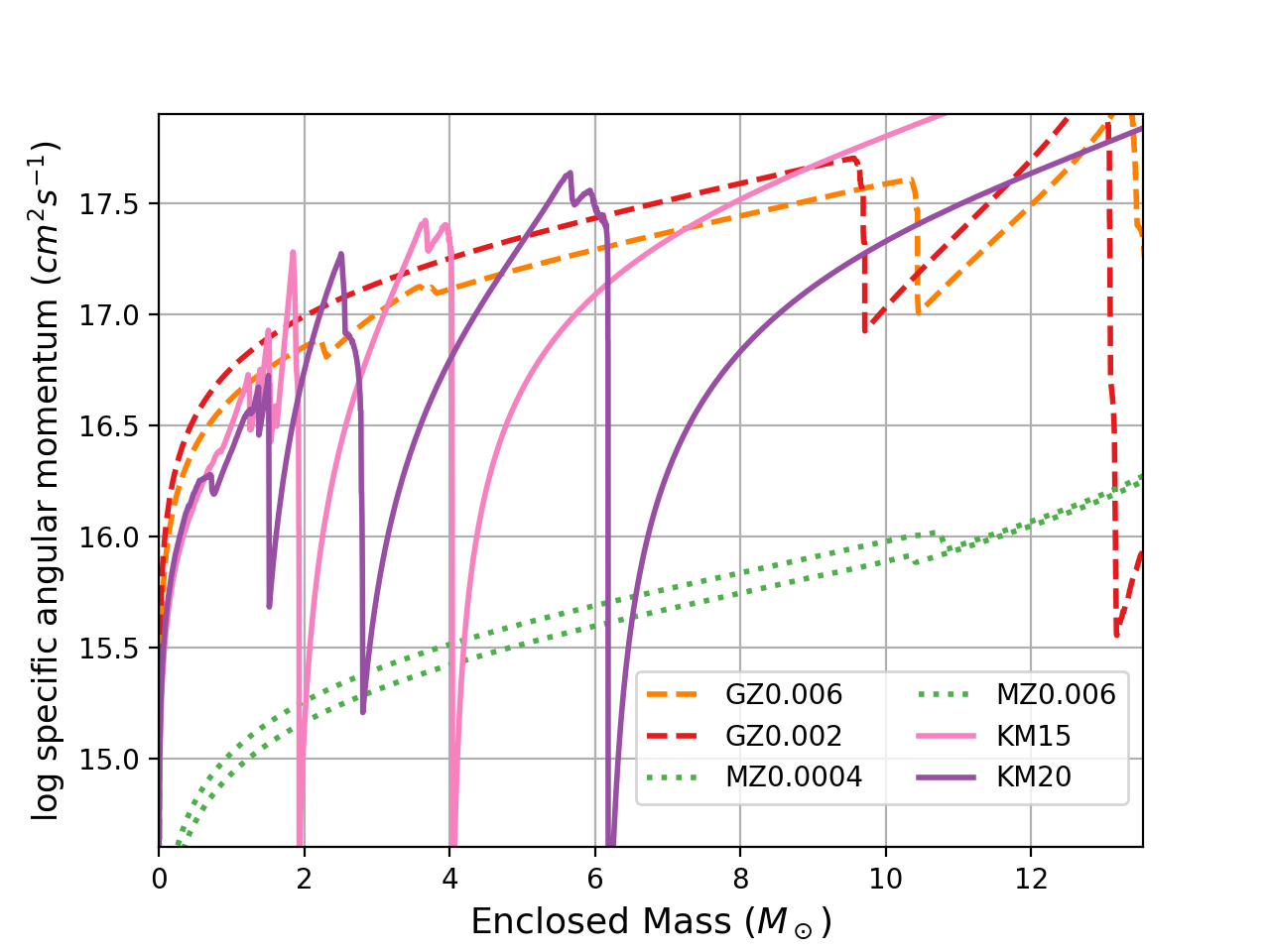}
    \caption{Angular momentum for 3 different mechanisms coupling the angular momenta between burning layers.  The {\it Genec} models (GZ0.006, GZ0.002) use no magnetic coupling between burning layers correspond to a 32\,M$_\odot$ star with metallicities set to 0.006 and 0.002.  Metallicity only mildly affects the spin rates for the 32\,M$_\odot$ star.  The {\it Kepler} models (KM15, KM20) using a mild Taylor-Spruit dynamo that reproduces pulsar spin velocities.  These models use 15 and 20\,M$_\odot$ solar metallicity progenitors.  The primary differences in the angular momentum profiles between the {\it Genec} and {\it Kepler} models lies in this burning layer coupling.  The {\it Kepler} models are evolved to collapse.  The {\it MESA} models (M0.006, M0.0004) were run with strong magnetic coupling between burning layers for a 40\,M$_\odot$ star.  This coupling highlights the strong dependence of the spin on the prescription for magnetic coupling.} 
    \label{fig:rot}
\end{figure}

The diversity in the rotation periods of current models of massive stars lies principally in the prescription of the coupling of angular momentum between burning layers.  Figure~\ref{fig:rot} shows the specific angular momentum for a range of different models, varying mass, metallicity and, most importantly, the method used to couple the different burning layers.  For single star models, the fastest-spinning cores are produced by the {\it GENEC}\citep{2008Ap&SS.316...43E} simulations that do not include strong coupling between burning layers.  The {\it KEPLER}\citep{2005ApJ...626..350H} models include a Taylor-Spruit dynamo coupling the stellar boundaries, producing slower rotating models.  Also shown are {\it MESA}~\citep{2013ApJS..208....4P} models using its version of the Taylor-Spruit dynamo with strong coupling from the high-magnetic field {\it MESA} models used in the \cite{2020A&A...636A.104B} paper.   The differences in the angular momentum are primarily due to the very different schemes used in these particular calculations to couple the different burning layers for these calculations and not in the codes themselves.  One other major difference is that the {\it KEPLER} models are modeled to collapse and the {\it GENEC} and {\it MESA} models are modeled to the onset of silicon burning.  

For this paper, we primarily focus on the different angular momentum profiles.  Typically, a specific angular momentum above $\sim 10^{17} {\rm \, cm^2 \, s^{-1}}$ is needed within the inner 1.4-2\,M$_\odot$ of the stellar core so that the NS spin energy equals $10^{52}\, {\rm erg}$ or a black hole accretion disk whose extent is 3 times that of the innermost stable circular orbit.  For black hole accretion disk systems, this can be achieved if the coupling is weak, but it is impossible for stars that are strongly coupled ({\it Genec} models - dotted lines).  But none of these models work for our magnetar engine.  We will discuss the ramifications of these results in Sections~\ref{sec:magnetar} and~\ref{sec:collapsar}.  

\begin{figure}
    \includegraphics[width=0.5\textwidth]{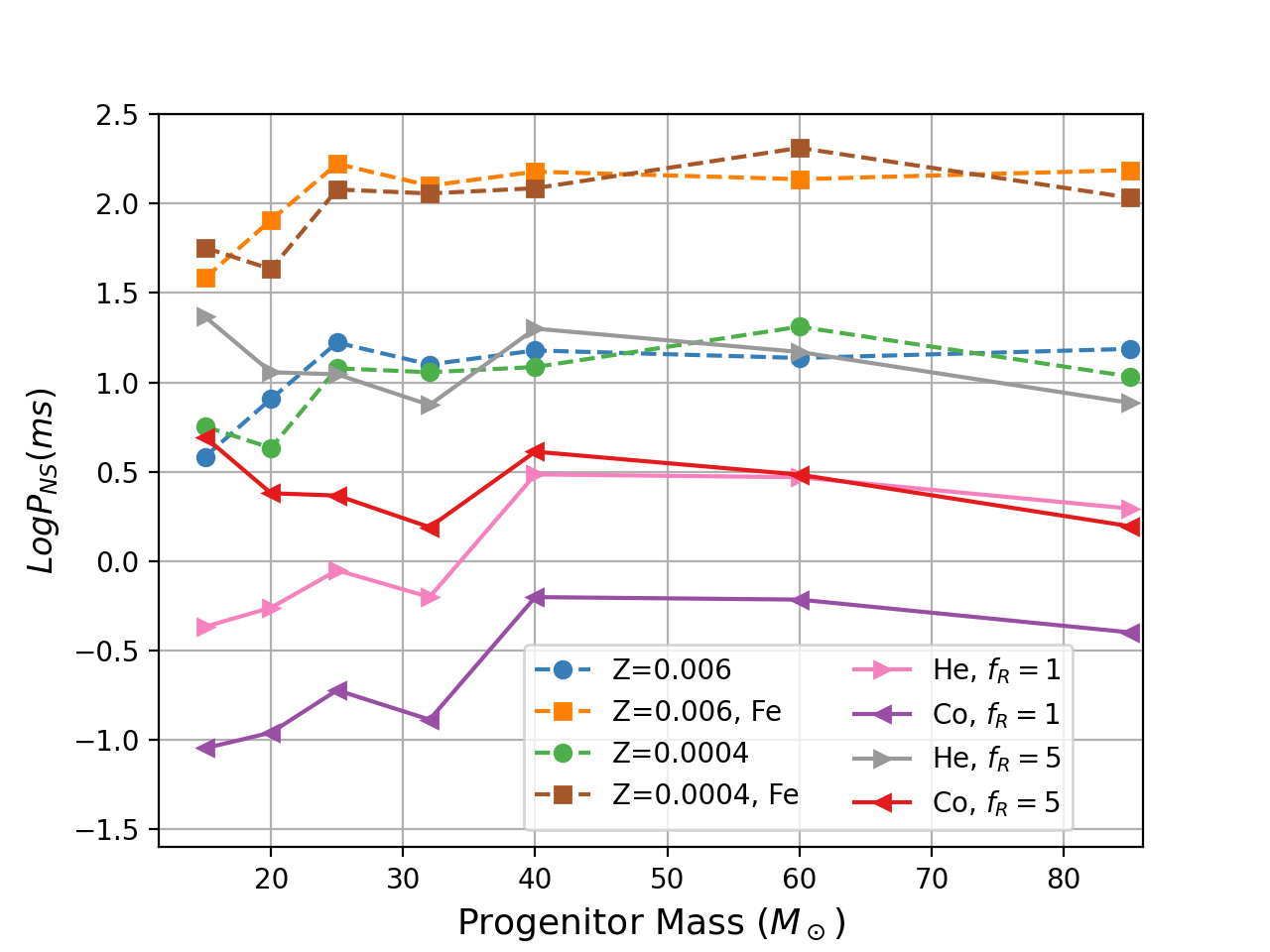}
    \caption{Spin periods of NSs formed in the collapse of massive stars.  Those stars where the proto-NS would collapse through continued accretion prior to cooling have spin periods calculated assuming a 30\,km proto-NS radius.  Progenitors that form NSs assume a final NS radius of 10\,km.  The Z=0.006, Z=0.0004 models correspond to different metallicities.  Models without the 'Fe' correspond to spin periods calculated assuming the angular momentum is set by the state prior to silicon burning and the 'Fe' models correspond to stars assuming mild coupling through the collapse of the iron core.  The He, CO models correspond to He and CO stars where the separation is set to a factor $f_R$ times the Roche radius.} 
    \label{fig:NSspin}
\end{figure}

From the angular momentum of these stars, we can calculate the spin of the NS or BH formed in the collapse by assuming the angular momentum in the star is preserved during the collapse and formation of the compact remnant.  However, this simple prescription would overestimate the total angular momentum.  If the angular momentum is sufficiently high that the material is centrifugally supported prior to its incorporation into the compact remnant, it will hang up in a disk.  This material must lose a fraction of its angular momentum to add its mass to the compact remnant.  This places an upper limit on the angular momentum accreted.  For NSs, this upper limit ($j_{\rm max}^{\rm NS}$) is:
\begin{equation}
    j_{\rm max}^{\rm NS} = \sqrt{G M_{\rm NS} r_{\rm NS}}
\end{equation}
where $G$ is the gravitational constant, $M_{\rm NS}$ is the compact remnant mass during the accretion and $r_{\rm NS}$ is the NS radius.  To estimate the NS spin, we collapse layer after layer of the star where the mass of the NS is:
\begin{equation}
    M_{\rm NS}^k = M_{\rm NS}^{k-1} + dm^k
\end{equation}
where the $k$ refers to the layer or zone from the stellar model and $dm^k$ is the mass of that zone.   The corresponding angular momentum of this accreting NS is:
\begin{equation}
    J_{\rm NS}^k = J_{\rm NS}^{k-1} + min(j^k,j_{\rm max}^{\rm NS}) dm^k
\end{equation}
where $j^k$ is the specific angular momentum in zone $k$.  During formation, the NS is hot and more extended than its final radius.  To get an upper limit on the angular momentum accreted for BH-forming systems, we assume $j_{\rm max}^{\rm NS}$ is set by this extended radius ($\sim 30\,{\rm km}$).  For systems that ultimately form a NS, we use a more compact 10\,km radius.  Figure~\ref{fig:NSspin} shows the expected spin periods for a set of massive stars, NS-forming progenitors as well as the spins of NSs in the initial collapse of BH-forming progenitors.  The single star models compare the results of the spins conserving the angular momentum profiles before and after silicon burning assuming weak Taylor-Spruit coupling as is used in most {\it Kepler} calculations (solid lines).  During silicon burning, angular momentum in the iron core is lost to the silicon layer, reducing the final NS spin period.  Since many stellar calculations only evolve the star to the onset of silicon burning, many spin estimates will be limited to these models which could overestimate the spin periods by an order of magnitude.  

Our single star models all used rapidly spinning progenitors (at formation, these stars are all within a factor of 2 of breakup spin velocities). For these single stars with Taylor-Spruit coupling, the fastest NS spin periods are all above a few ms.  These models are a good fit for the fastest spinning pulsars~\citep{2006ApJ...643..332F,2013MNRAS.430.2281N}.   The fastest spins are produced by stars that collapse to form NSs.   But, as we shall discuss below, these stars struggle to produce the high energies seen in GRBs.   Any observations of a $\sim 1 \, {\rm ms}$ pulsar is either an indication that the angular momentum coupling is not as strong as the Taylor-Spruit dynamo predicts or that tidal spin-up has occurred.   It is possible to gain angular momentum in the explosion itself from non-rotating stars.  Simulations have produced spins anywhere between 10\,ms to many seconds from a non- or slowly rotating star~\citep{2007Natur.445...58B,2007ApJ...659.1438F,2011ApJ...732...57R,2013A&A...552A.126W,2024ApJ...963...63B}.  As we shall see in Section~\ref{sec:magnetar}, more rapid spins are required for the magnetar engine to produce lGRBs.  

Hereafter, we focus on tight binaries that undergo tidal spin-up.  Common envelope mass loss can produce a wide range of helium star binaries with either a main sequence or compact object companion.  Further removing the helium envelope through binary mass transfer will require very tight binaries, most likely with compact companions.  In Figure~\ref{fig:NSspin}, we assume only the tightest binaries, assuming a compact binary with an orbital separation set to a factor, $f_R$, of the Roche radius.  For these tight, tidally spun-up stars, the spin periods can fall below 1\,ms.  As we shall discuss in Section~\ref{sec:magnetar}, such fast spins will lead to instabilities that remove angular momentum from the system.  

\begin{figure}
    \includegraphics[width=0.5\textwidth]{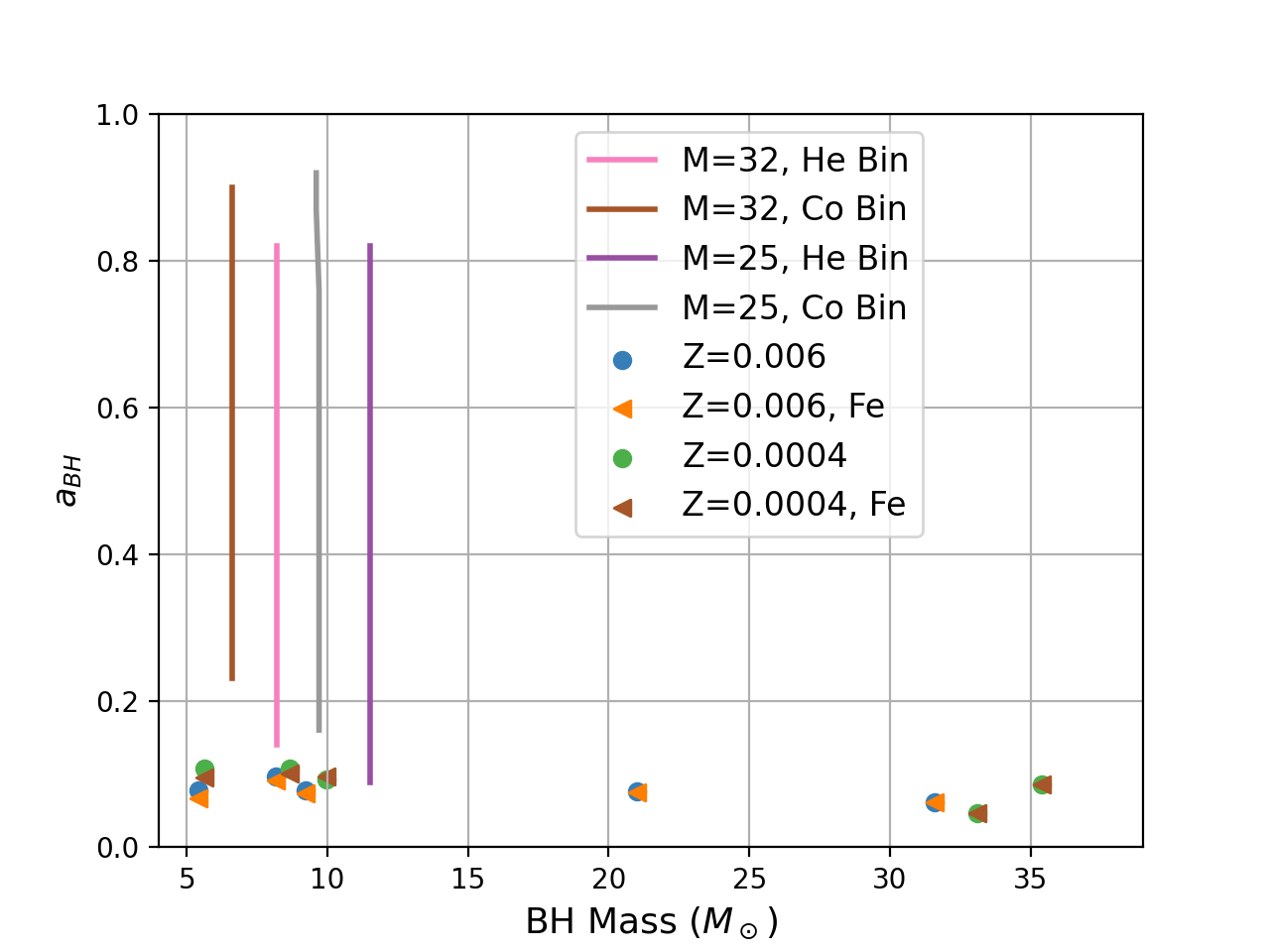}
    \caption{Black hole spins formed from our stellar collapse models.  Here we limit the models to black-hole forming stars and estimate the mass from the rapid models of ~\cite{2012ApJ...749...91F}.  The single-star and binary models considered are the same as those in Figure~\ref{fig:NSspin} where the lines for the binary models correspond to separations ranging from 1-5 times the Roche radius.} 
    \label{fig:BHspin}
\end{figure}

Conservation of angular momentum can also be used to estimate the mass and spin of the black hole.  We set the maximum angular momentum accreted onto the black hole ($j_{\rm max}^{\rm BH})$ to:
\begin{equation}
    j_{\rm max}^{\rm BH} = \sqrt{G M_{\rm BH} r_{\rm EH}}
\end{equation}
where $M_{\rm BH}$ is the black hole mass and where we assume the angular momentum is lost until the matter reaches the event horizon:
\begin{equation}
    r_{\rm EH}/r_{\rm Schwarzschild} = 1 + \sqrt{1-a^2 cos(\theta)^2}
\end{equation}
where, because we are interested in the plane of the angular momentum, $\theta=\pi/2$ and we use  
\begin{equation}
    r_{\rm Schwarschild} = 2 G M_{\rm BH} / c^2
\end{equation}
where $c$ is the speed of light.  But disk formation can also alter the mass accretion.  In disk models, a non-negligible amount of mass is ejected along with the angular momentum.  These disk winds can eject anywhere from 1-30\% of the disk mass.  Ultimately, the energy injected into the star from the GRB jet and the accretion disk wind will prevent further accretion.  This process is not well understood and, for our initial black hole mass and spin estimates, we will only include the mass loss from the disk wind.  The resulting black hole masses and spins are:
\begin{equation}
    M_{\rm BH}^k = M_{\rm BH}^{k-1} + dm^k (1-f_{\rm wind})
\end{equation}
and
\begin{equation}
    J_{\rm BH}^k = J_{\rm BH}^{k-1} + min(j^k,j_{\rm max}^{\rm BH}) dm^k.
\end{equation}
The resulting spins for black holes using our models are in Figure~\ref{fig:BHspin}. 

\begin{figure}
    \includegraphics[width=0.5\textwidth]{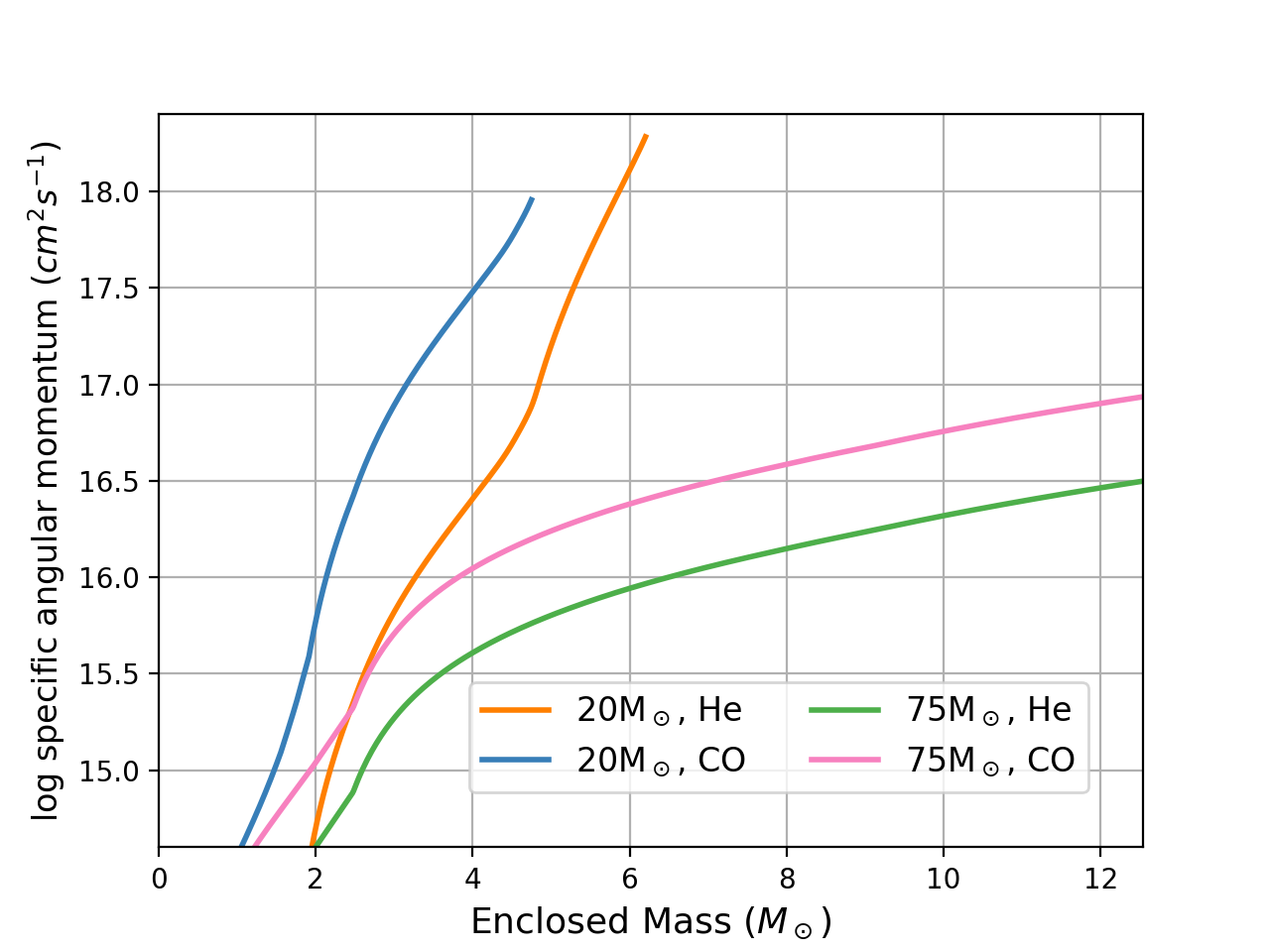}
    \caption{Angular momentum versus enclosed mass for 4 tidally-locked binaries assuming separations near the Roche radius for CO and He stars with zero-age main-sequence progenitor masses of 20, 75\,M$_\odot$.  The angular momentum for all these models exceeds $10^{17} \, {\rm cm^2 \, s^{-1}}$ for all our models, but he 20\,M$_\odot$ produces extremely high angular momenta.} 
    \label{fig:rotbin}
\end{figure}

For single-stars using the Spruit-Taylor dynamo, black hole spins tend to be 0.1.  Such models are consistent with the observed spins from gravitational wave binaries~\citep{2020A&A...636A.104B}.  But a subset of BHs in these binaries appear to be spinning more rapidly.  One explanation for for the progenitors of such BHs is that they are spun up through tidal forces in a close binary.  The angular momentum profiles for these binaries lie somewhere between the ``no coupling'' {\it GENEC} calculations and the more strongly coupled {\it MESA} models (Figure~\ref{fig:rotbin}).  The closer binaries produced in CO core binaries lead to higher angular momenta at any given mass coordinate, but the helium stars are more extended and the outermost layers can have specific angular momenta that exceed that of the CO stars.

\begin{figure}
    \includegraphics[width=0.5\textwidth]{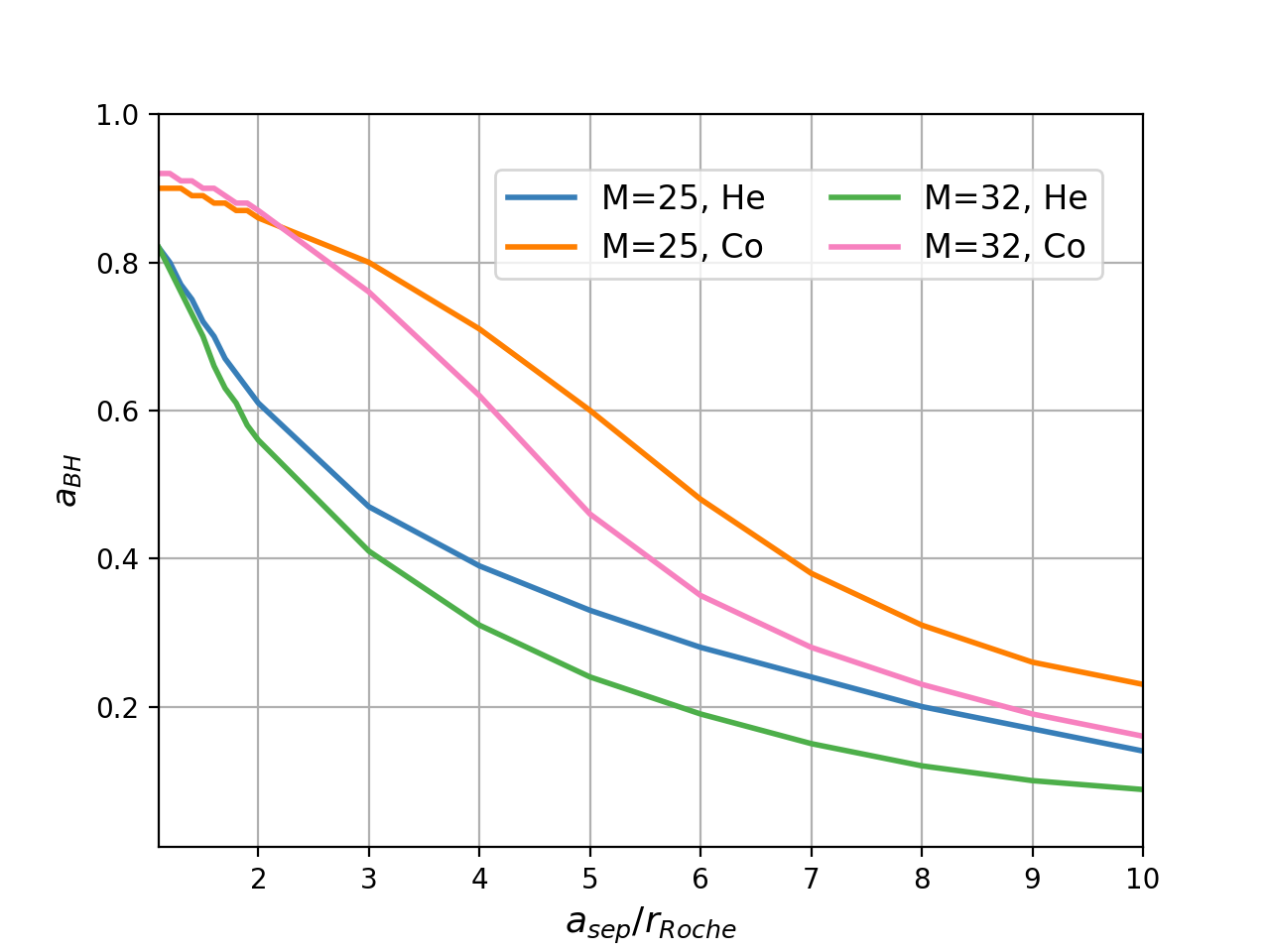}
    \caption{Black Hole dimensionless spin parameter versus of orbital separation (in Roche radii) for 4 models assuming He and CO stars for 25, 32\,M$_\odot$ zero-age main-sequence mass progenitors.  Spin rates above 0.4 are expected for models that produce reasonable disks.} 
    \label{fig:BHSpinvsa}
\end{figure}

The angular momentum in these binary systems depends upon how close the binaries become.  Figure~\ref{fig:rotbin} assumed tight binaries just above the Roche separation where mass transfer would occur.  The lines in Figure~\ref{fig:BHspin} show the expected spin parameters for orbital separations for these black holes ranging from 1.1-5 Roche radii.  The separation of the binary is set by when the common envelope phase ejects the envelope.  For He-star forming binaries, the initial separation can be anywhere where the hydrogen giant phase can envelope the companion.  These common envelope systems can form a wide range of separations, leading to systems that form black holes across (and beyond) the range shown in Figure~\ref{fig:BHSpinvsa}.  But helium common envelope or other binary mass ejection mechanisms are likely to only occur in very tight systems (common envelope requires the expansion of the star and helium stars do not expand much in most stellar evolution calculations), so such systems will start in tight binaries and only become tighter (separations within a few Roche radii).  

Carbon/Oxygen stars, the progenitors of Type Ic binaries, are more compact than He-stars and, hence, can produce tighter binaries.  Hence, the BH spins are higher for these CO stars.  Hydrogen common envelope scenarios will produce a wide range of separations between the helium core and its companion star.  Only a small fraction of these common envelope inspirals will produce the tight binaries needed to produce fast rotating cores.  This is because hydrogen envelopes are extended and easily ejected in a common envelope.  Most of the helium stars in these systems will not be spun up through tidal locking.  CO cores are in a different situation.  Helium stars are less extended.  If a helium binary mass ejection occurs, the binary will be tight.  The question for this scenario is whether helium binary mass ejections even occur (CO cores can be produced simply through mass loss in strong Wolf-Rayet winds).  For our discussion, we will assume such helium common envelopes can occur.

\section{Magnetar Energies}
\label{sec:magnetar}

Magnetar engines tap the rotational energy in the newly-formed NSs to power emission or a jet.  With the spin periods from Section~\ref{sec:spins}, we can calculate this energy.  The moment of inertia for neutron stars depends upon the equation of state~\citep{2008ApJ...685..390W}, but all estimates of the moment for a NS ($I_{\rm NS}$) are within a factor of two of:
\begin{equation}
I_{\rm NS} = 10^{45} (M_{\rm NS}/M_\odot) {\rm~g~cm^2}
\end{equation}
where $M_{\rm NS}$ is the NS mass. The corresponding
rotational energy ($E_{\rm rot}$) is:
\begin{equation}
    E_{\rm rot} = 5 \times 10^{50} (\omega/1000\,{\rm Hz})^2 \rm{erg}
\end{equation}
where $\omega$ is the angular velocity.  A NS with a spin period of 1\,ms has a total of $2 \times 10^{52} {\rm \, erg}$.  

The maximum spin period of NSs is limited by instabilities in the NS.  The onset of these instabilities occurs when the rotational energy exceeds 14\% of the potential energy of the neutron star~\citep{1983bhwd.book.....S}:
\begin{equation}
\beta = E_{\rm rot}/|W| > 0.14
\end{equation}
where 
\begin{equation}
    |W| = G M_{\rm NS}^2/r_{\rm NS} \approx 5 \times 10^{53}\,{\rm erg}
\end{equation}
and, for the neutron star mass and radius, we have $M_{\rm NS}=1.4\,M_\odot, r_{\rm NS} = 10\,{\rm km}$, respectively.  These instabilities place an upper limit on the energy available for a magnetar of $\approx 7 \times 10^{52} {\rm erg}$.  But other instabilities can further reduce the maximum total energy.  For example, at extremely high spin rates, Rossby waves can develop, driving the total rotational energy down another order of magnitude~\citep{2000ApJ...543..386H,2001ApJ...549.1111L}.  At a fixed angular momentum, the rotational energy in the proto-neutron star increases as it becomes more compact.  If the magnetar-strength magnetic fields develop when the proto-neutron star is still hot and extended, it can lose its angular momentum before the rotational energy reaches its peak.  For a given amount of total angular momentum ($J_{\rm tot}$), the total rotational energy is inversely proportional to the square of the radius:
\begin{equation}
    E_{\rm rot} = 1/2 I_{\rm NS} \omega^2 = 1/2 J_{\rm tot}^2/I_{\rm NS} \propto J_{\rm tot}^2/r_{\rm NS}^2.
\end{equation}
A hot NS has a radius roughly 3 times that of a cold NS.  At this point, the total rotational energy is roughly 10 times lower than it will have when it cools.  For BH forming systems, this is the only magnetar energy reservoir, making it difficult for magnetar engines to have much power for these systems.  The most powerful magnetars are likely to be produced in NS-forming systems where the magnetar-like fields are not completely formed until the NS cools.

\begin{figure}
    \includegraphics[width=0.5\textwidth]{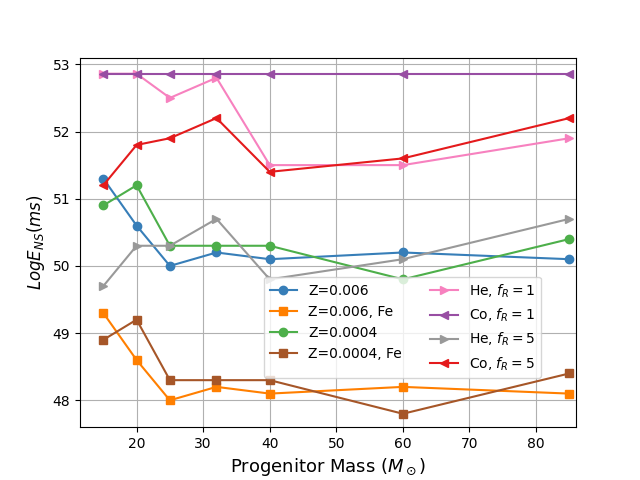}
    \caption{Available rotational energy for neutron stars in stellar collapse as a function of zero-age main sequence mass.  Because the neutron star only exists in its hot, extended state for the BH-forming, progenitor masses above 25\,M$_\odot$ can only have weak magnetars.} 
    \label{fig:mage}
\end{figure}

With the upper limit set by NS instabilities (ignoring Rossby waves and allowing the NS to cool where possible), we can estimate the rotational energy available for magnetar engines.  Figure~\ref{fig:mage} shows the range of energies for both single stars and binary systems.  For single stars, the rotational energies available for magnetar engines lie below $\sim 10^{51} {\rm erg}$.  For binary systems, particularly close CO stars, this rotational energy reaches the upper limit available.

From these results, we can identify trends in lGRBs and their associated SNe.  For our single star models assuming coupling caused by the Tayler-Spruit dynamo~\citep{2002A&A...381..923S}, the magnetar energies are typically below $10^{50} {\rm \, erg}$.  The most powerful magnetars arise from binaries.  A broad range of CO binaries achieve rotational energies near the maximum value derived above.  Only the tightest He star binaries will have the energies to produce normal GRBs.  Under these progenitors, we would expect most GRB-associated SNe to be classified as type Ic.  However, this progenitor also argues that the most energetic magnetars should occur from lower-mass progenitor stars and we'd expect no significant metallicity or redshift dependence under this engine.  The only effect of metallicity would be the fact that the initial mass function flattens with lower metallicty.  Under this engine, the rate of GRBs as a function of star formation, should decrease with decreasing metallicity and increasing redshift (see Section \ref{sec:metallicity}).

\section{BHAD Accretion and Energies}
\label{sec:collapsar}

BHAD engines rely, first and foremost, upon the formation of an accretion disk.  Models with modest coupling (e.g. our {\it Kepler} models) are just at the boundary between insufficient and sufficient angular momentum to produce a disk.  A number of mechanisms have been discussed to extract energy from accreting BHs~\citep{2019EPJA...55..132F}, tapping energy either from the energy in the disk or rotational energy of the BH.  But all of these require an accretion disk either as an energy source or as a mediator.  To understand BH accretion disk engines, we must first understand the accretion rate and accretion disk evolution.

Accretion rates and durations provide insight into the properties of lGRBs.  The timescale of disk accretion is set by the duration at which the disk is fed by the collapsing star and the accretion timescale of the disk.  The disk feeding timescale is often approximated by the free-fall time of material collapsing onto the compact object~\citep{2019EPJA...55..132F}.  This free-fall time ($t_{\rm ff}$) is set by the enclosed mass within that radius corresponding to the BH mass ($M_{\rm BH}$) and the position of the free-falling material ($r$) at the time of collapse:
\begin{equation}
    t_{\rm ff} = \pi r^{3/2}/\sqrt{8 G M_{\rm BH}}.
\end{equation}
This timescale gives the approximate time at which the material forms in a disk.

\begin{figure}
    \includegraphics[width=0.5\textwidth]{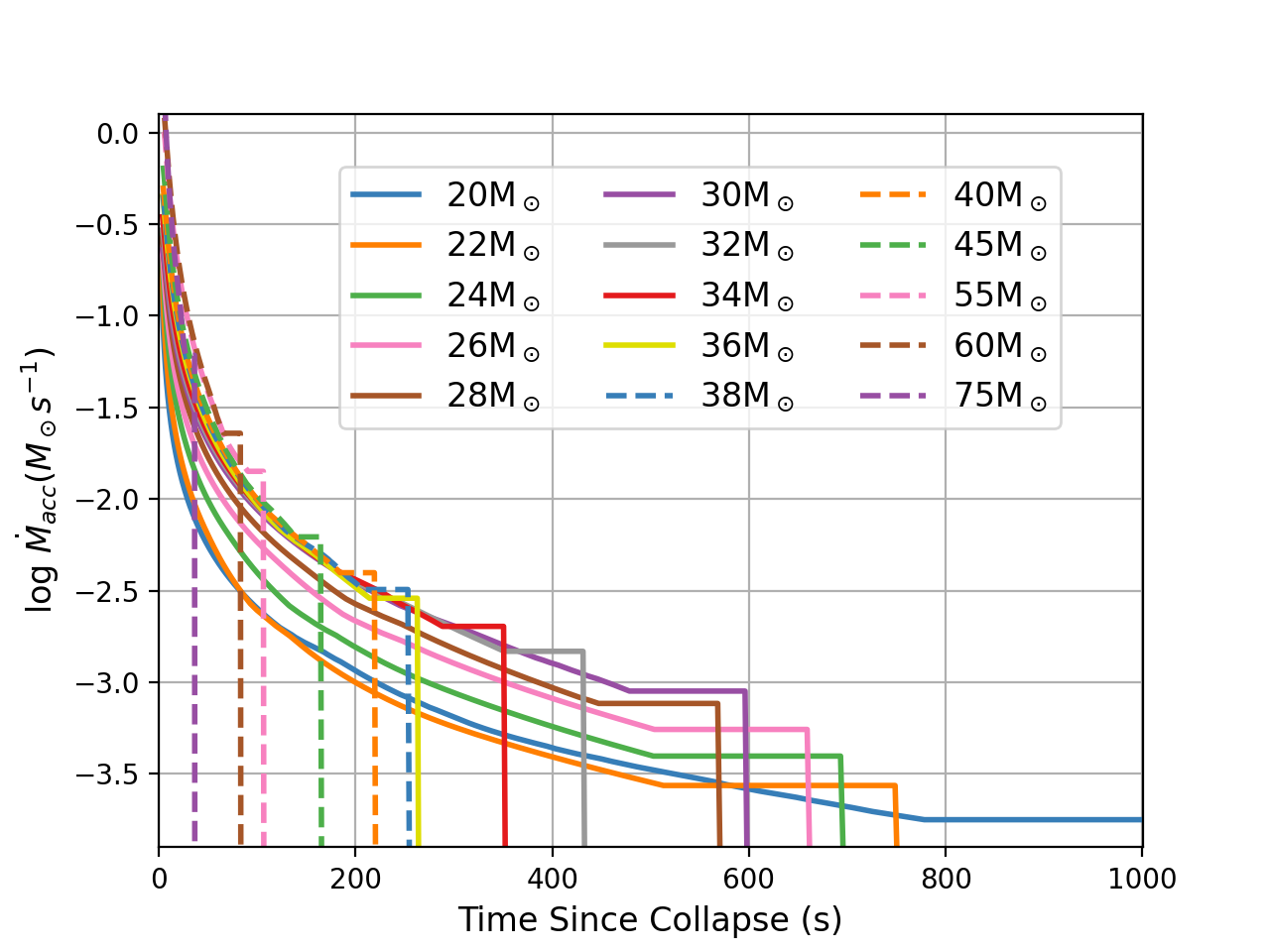}
    \caption{Disk accretion rate as a function of time for tight-binary ($f_R = 1.0$) systems with zero age main-sequence masses ranging from 20-75\,M$_\odot$.  The more common, 20\,M$_\odot$ progenitors continue to accrete for over 1000\,s.} 
    \label{fig:mdot}
\end{figure}

The disk accretion timescale can be estimated from an $\alpha$-disk model~\citep{1999ApJ...518..356P}.  In the alpha-disk prescription, the accretion timescale ($t_{\rm acc}$) of the disk is set to the orbital period ($P_{\rm disk}$) of matter in the disk divided by an efficiency parameter ($\alpha$):
\begin{equation}
    t_{\rm acc} = P_{\rm disk}/\alpha = 2 \pi r_{\rm disk}^{3/2}/(\alpha \sqrt{G M_{\rm BH}})
\end{equation}
where $r_{\rm disk}$ is the radial extent of the disk set by the specific angular momentum, $j(r)$, in the star:
\begin{equation}
    r_{\rm disk} = j(r)^2/(G M_{\rm BH})
\end{equation}
and the corresponding accretion time through the disk in terms of angular momentum is:
\begin{eqnarray}
     t_{\rm acc} & & = 2 \pi j^3(r) /\alpha/(G M_{\rm BH}^2) = \\ \nonumber
    & & 3.9s \left(\frac{j(r)}{10^{17} {\rm cm}^2 {\rm s}^{-1}}\right)^3 \left(\frac{0.01}{\alpha}\right) \left(\frac{3M_\odot}{M_{\rm BH}}\right)^2
\end{eqnarray}
For our binary models, we assume the duration of the jet is the combination of the free-fall and accretion timescales.  For the fastest rotating models, this time is set by the disk accretion timescale, but for the more compact cores, the timescale is close to the free-fall time.  The corresponding accretion rates using these accretion parameters for a range of Kepler pre-collapse progenitors~\citep{2002RvMP...74.1015W} are shown in Figure~\ref{fig:mdot}.

\begin{figure}
    \includegraphics[width=0.5\textwidth]{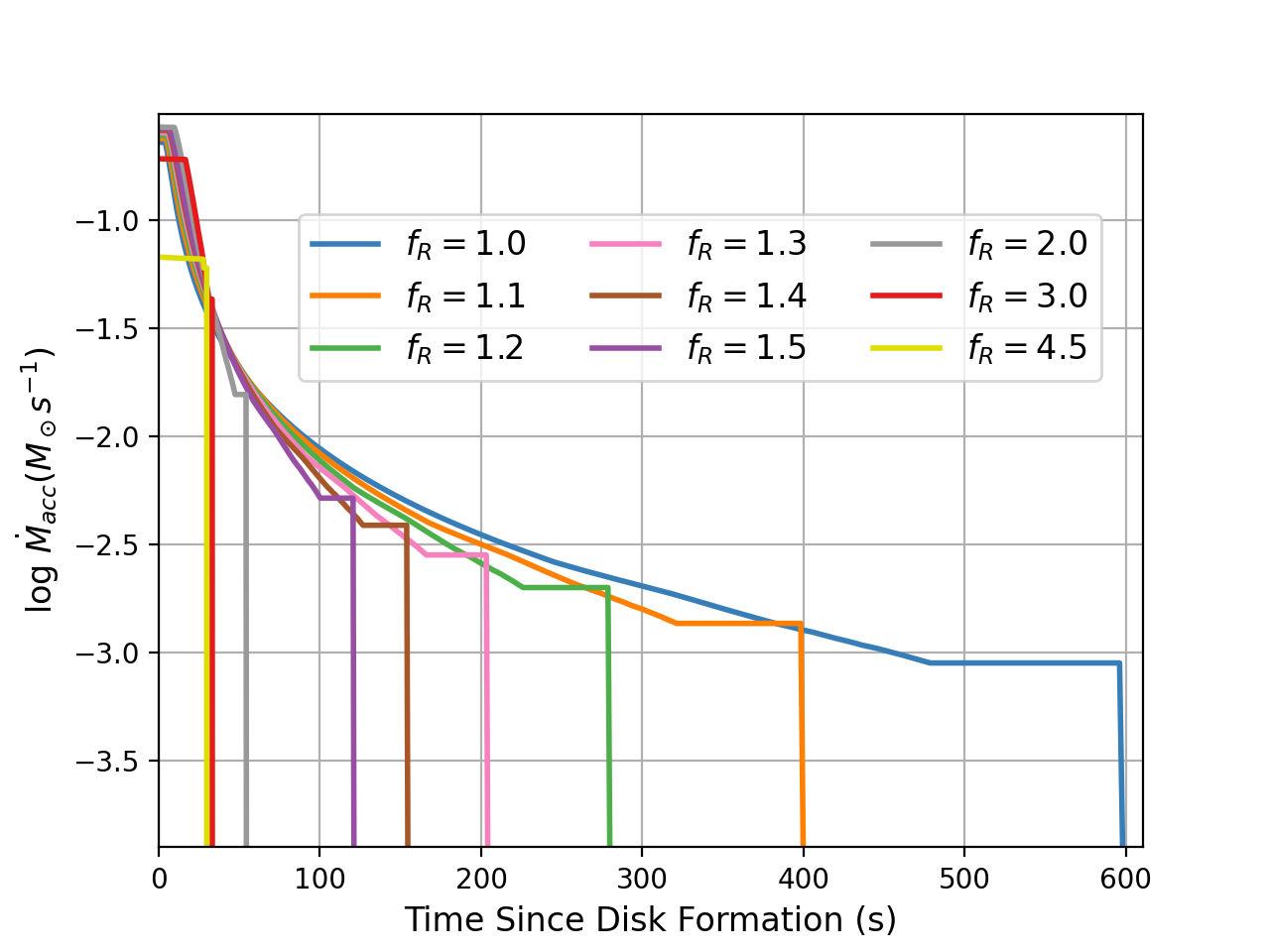}
    \includegraphics[width=0.5\textwidth]{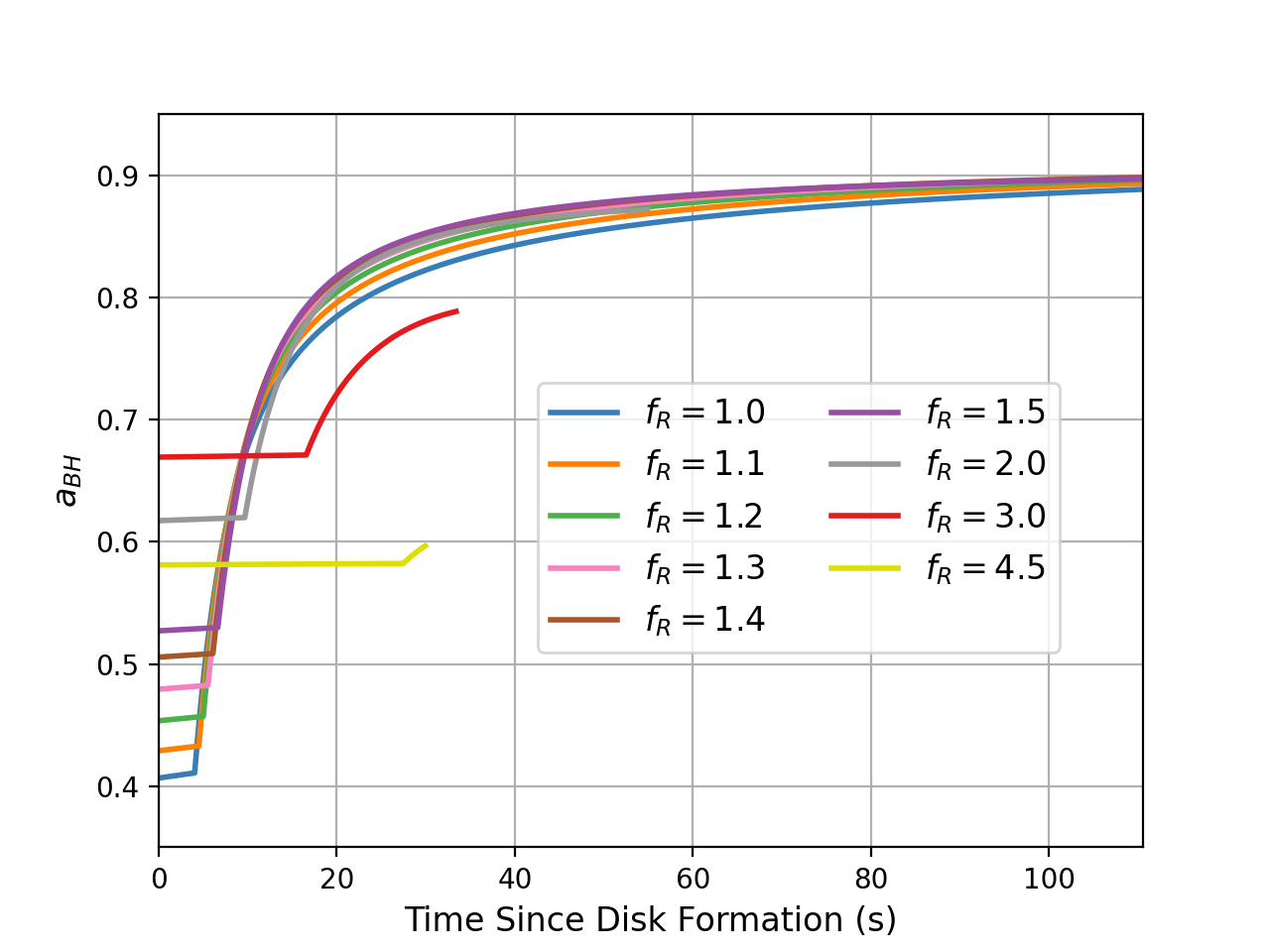}
    \includegraphics[width=0.5\textwidth]{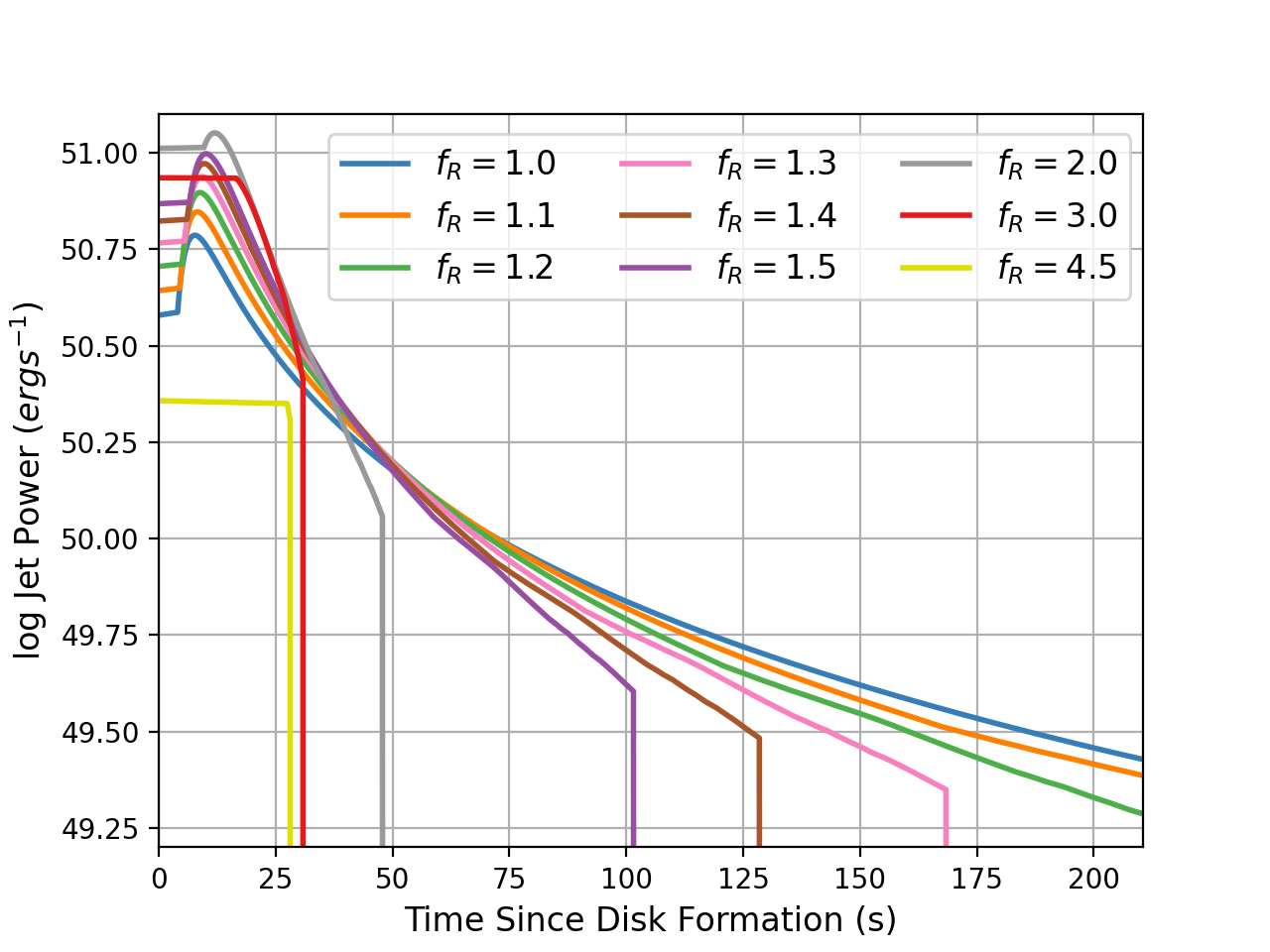}
    \caption{Top panel:  Accretion rate versus time after the formation of the disk for a 30\,M$_\odot$ zero-age main-sequence progenitor as a function of time after the formation of the disk.  Disk formation occurs later for wider binaries.   Middle panel:  Corresponding black hole spin rate as a function of this time.  Bottom Panel:  GRB jet power as a function of time using Equation~\ref{eq:BZform}.} 
    \label{fig:lbz}
\end{figure}

For our smallest BH-forming systems, the accretion rate can extend out to 1000\,s, potentially explaining the ultra-long GRBs (see Section~\ref{sec:GRB}).  The accretion timescale shortens for wider binaries.  The top panel of Figure~\ref{fig:lbz} shows the accretion rate as a function of time for binaries at different separations.  As the binary becomes wider, the time to form a disk gets longer.

It is possible that the jet-driven explosion will disrupt the star~\citep{1999ApJ...524..262M}, shortening the accretion duration.  But there are also methods to extend the accretion timescale.  One way to extend the accretion phase, and hence drive the engine longer, is to argue that the accretion process is much more complex than the $\alpha$-disk model suggests.  One such process is the magnetically arrested disk (MAD) where a poloidal field produced in the collapse or wound up in the disk creates enough pressure at the event horizon to halt accretion~\citep[for an overview, see][]{2016MNRAS.461.1045L}.   But MAD disks are unlikely to extend the duration of the accretion disk considerably.  Alternative mechanisms have been proposed to extend the emission relying on dense or baryon-loaded ejecta~\citep{2014MNRAS.445.2414V,2015ApJ...806..205D}.  Another way to extend the disk accretion timescale is to tap the angular momentum in the BH to prop up the disk~\citep{1999Sci...284..115V}.  We will discuss the role of disk durations and their impact on GRB signals in Section~\ref{sec:GRB}.

For many BHAD mechanisms, the power of the accretion disk is set by the kinetic energy in the disk which, in turn, depends upon the structure of the disk (accretion rate) and BH spin.  \cite{1999ApJ...518..356P} calculated these disk properties, providing an estimate of the jet energy as a function of BH spin and accretion rate.   These results assumed that the strength of the magnetic field was limited by the energy in the disk (i.e. the magnetic field energy density could not exceed the disk thermal energy density).  Fitting to these models, \cite{2003ApJ...591..288H} developed a formula for the jet power ($L_{\rm jet}$):
\begin{eqnarray}
    L_{\rm jet} & = & \frac{f_{\rm jet}}{0.01} 10^{50} 10^{0.1/(1-a_{\rm BH}) - 0.1} \\ \nonumber
    & & \left(\frac{3M_\odot}{M_{\rm BH}}\right)^3 
    \frac{\dot{M}_{\rm disk}}{0.1\,M_\odot s^{-1}} {\rm erg \, s^{-1}}
\end{eqnarray}
where $\dot{M}_{\rm disk}$ is the accretion rate as shown in Figure~\ref{fig:mdot} and $f_{\rm jet}$ is a factor roughly describing the efficiency.  Because the energy in the disk decreases with larger BH radii, this mechanism decreases with BH mass. 

The original \cite{1977MNRAS.179..433B} predicts a jet power that is much less sensistive to the disk energy density: 
\begin{equation}
    L_{\rm jet} = 3 \times 10^{51} a^2_{\rm BH} \frac{\dot{M}_{\rm disk}}{0.1\,M_\odot s^{-1}} {\rm erg \, s^{-1}}
    \label{eq:BZform}
\end{equation}
Other studies follow this same trend, predicting extremely powerful jet engines.  \cite{2023ApJ...952L..32G} argue that the jet power is:
\begin{equation}
    L_{\rm jet} = 2 \times 10^{53} (1.063 a^4_{\rm BH} + 0.395 a^2_{\rm BH}) \frac{\dot{M}_{\rm disk}}{0.1\,M_\odot s^{-1}} {\rm erg \, s^{-1}}
\end{equation}
For this latter jet prescription, we can achieve high energies even for slow rotating BHs.  The angular momentum requirement for this engine is not set by the energy requirements of the GRB, but the angular momentum required to produce a disk.  

We use Equation~\ref{eq:BZform} to determine the power in the jet in Figure~\ref{fig:lbz}.  For the wider binaries, the disk forms later, but the BH has accreted more so the spins are, if anything, higher.  The final angular momentum does not reach the high values ($a_{\rm BH}>0.9$) that our tight binaries achieve.  The resulting power in these systems is weaker and has a much shorter duration.  Our short duration disks can either form from massive progenitors (producing strong explosions) or wide binaries (producing weak explosions).  Although our ultra-long durations can have initially strong GRB jet-power, the late-time power should be much less.  

The jet drives a shock through the star and this shock can drive the explosion of the star~\citep{1999ApJ...524..262M}.  Winds from the disk drive an outflow that can also contribute to the explosion energy.  The fraction of the disk mass ejected in this wind can be anywhere from 1-30\% of the disk mass and the velocities are within a factor of a few of the escape velocity.  Depending on the properties of the disk, the disk wind ejecta can carry considerable energy (more than $10^{51} \, {\rm erg}$) and the SNe produced in these massive-star GRBs are driven by a combination of this disk-wind and jet energies.  The disk winds will dominate the nucleosynthetic yield produced during the explosion itself (we will discuss this in Section~\ref{sec:nuc}).

\section{Observational Features}
\label{sec:theoryobs}

The specific progenitors for highly-rotating magnetar, NSAD and BHAD models will have observational features that can allow us to distinguish between these strong SN explosions and normal SNe.  With more detailed studies, we may be able to distinguish between these different high-spin engines.  Here we discuss a set of potentially distinguishing features of these engines.

\subsection{Nucleosynthetic Yield Signatures}
\label{sec:nuc}

The mechanisms producing the nucleosynthetic yields from core-collapse stars with convective-driven explosions can be categorized into three components: 1) isotopes produced in burning layers during stellar evolution, 2) isotopes produced in the convective region (typically iron peak elements), and 3) isotopes produced in the explosive shocks (primarily producing a range of alpha-chain elements).  Similarly, models invoking accretion disks also have three production components:  1) isotopes produced in burning layers during stellar evolution, 2) isotopes produced in the accretion disk and ejected in a viscosity-driven wind (typically iron peak elements), and 3) isotopes produced as the wind ejecta plows through the star.

A few general comments guide the nature of the yields in these explosions.  The production of r-process elements requires the collapsing stellar material to become neutron-rich.  Although electron capture can occur both in the convective and disk-driven engines, neutrinos from the NS tend to reset this neutronization and, for most models, these explosions do not produce elements beyond the iron peak or the first r-process peak.  Even for BH forming core-collapse stars, the densities and temperatures in the disk are typically not sufficiently high to drive significant deleptonization~\citep{1999ApJ...518..356P,2006ApJ...643.1057S} to produce significant amounts of heavy r-process and recent results arguing that r-process production in these events are typically hampered by numerical artifacts~\citep{2020ApJ...902...66M}.  The amount of deleptonization is easily estimated as a function of the accretion rate.  Figure~\ref{fig:mdot} shows the disk accretion rates as a function of progenitor mass and progenitor rotation.  Based on the Popham et al. disk models and their predicted neutrino luminosities~\citep{1999ApJ...518..356P}, we can estimate the deleptonization within the disk:  
\begin{equation}
    dY_e/dt \approx N_{\rm p}/(L_{\nu_e}/\epsilon_{\nu e}) t_{\rm acc} 
\end{equation}
where $N_{\rm p}$ is the number of protons, $L_{\nu_e}$ is the neutrino luminosity from electron capture, $\epsilon_{\nu e}$ is the neutrino energy and $t_{\rm acc}$ is the accretion timescale.  The electron capture luminosity is highly sensitive to the temperatures in the disk ($L_{\nu_e} \propto T^6$ where $T$ is the temperature).  Because of this, accretion rates above 0.5-1M$_\odot {\rm \, s^{-1}}$ are needed to reduce the electron fraction below 0.4.  As such, strong deleptonization is limited to NS mergers~\citep[e.g.][]{2019PhRvD.100b3008M} and our most massive stars (low metallicity stars above $\sim 60\,M_\odot$ see Figure~\ref{fig:mdot}).  

\begin{figure}
    \includegraphics[width=0.5\textwidth]{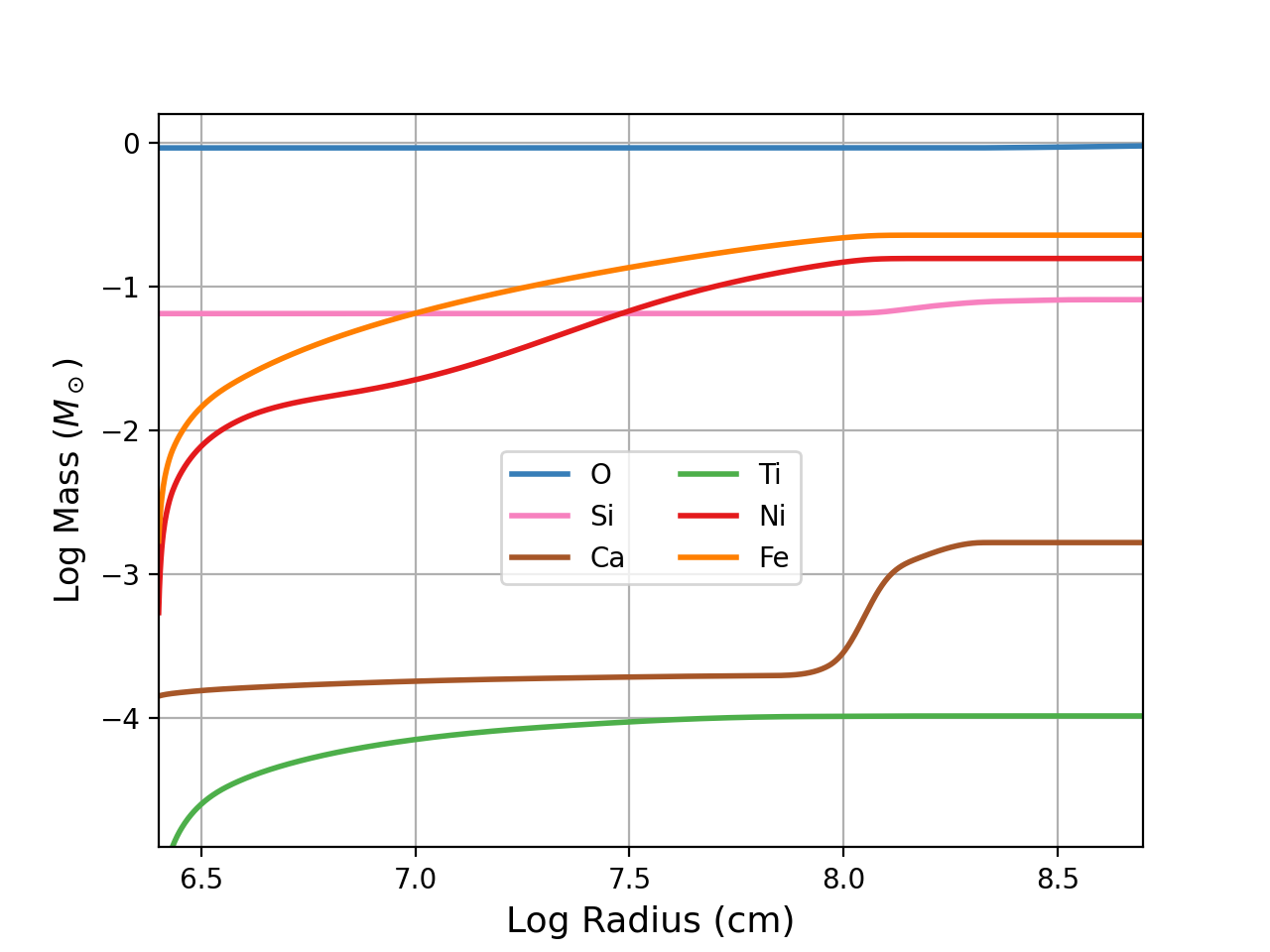}
    \caption{Abundances from a disk wind as a function of the extent of the disk.  This assumes that the wind ejects matter along most of the disk (using the model from \cite{2023ApJ...956...71K}.} 
    \label{fig:abundisk}
\end{figure}

Because of this, we will focus our nucleosynthetic yield study on the production of alpha elements up to the iron peak from our disk yields.  In these disk-driven explosions, disk winds dominate the production of heavy (iron peak) elements.  And the production of these elements depends sensitively on the maximum radius of the disk (more compact disks eject less mass).  Using the low entropy disk models from \cite{2023ApJ...956...71K}, we calculate the disk yields and combine it with the ejecta from stellar evolution.  The total yields as a function of the extent of the accretion disk is shown in Figure~\ref{fig:abundisk}.  The $^{56}$Ni yield, one of the power sources for the SN transients from these explosions, varies by an order of magnitude depending up the disk radius.  If this is the only energy source, the associated supernova luminosities from these explosions can range over an order of magnitude.  However, as we shall discuss below, these transients can also be powered by shock heating.

\subsection{Circumstellar Medium}

The circumstellar medium for our tidally-locked binaries is determined by the mass transfer events that produce the tight binaries in our magnetar, NSAD and BHAD progenitors.  This explosive mass-loss will produce heterogeneities in the circumstellar medium that may be observed in the resulting transient emission.  Shock interactions with the ejecta from the last mass transfer phase, the removal of the helium envelope, are likely to have the most dramatic effect on the transient emission.  Here we estimate the properties of this mass loss based on stellar evolution models.

We estimate the ejecta mass by assuming that the final mass transfer phase occurs somewhere between the end of core helium burning to the beginning of carbon ignition.  For this study, we use the time-scales and helium-shell masses inferred stellar evolution models reviewed in \cite{2002RvMP...74.1015W}.  Table~\ref{tab:csm} shows the range of time prior to collapse of the end of helium burning (He depletion) and carbon ignition.  The timescales differ for NS and BH forming stars.  The more massive stars that form BH systems have shorter timescales and more massive ejecta masses than those forming NSs.

\begin{table*}
\begin{center}
\begin{tabular}{l|ccccccc}
\hline\hline              
Engine & Time$_{\rm He Dep}$ & Time$_{\rm C Ign}$ & R$_{\rm shell, He Dep}$ &  R$_{\rm shell, C Ign}$ & T$_{\rm interaction}^{HeDep}$ & T$_{\rm interaction}^{C Ign}$ &M$_{\rm ejecta}$ \\
\hline
BHAD  & $\sim 10^4$y & 1-10y & $\sim 10$\,pc & 0.001-0.01\,pc & 200y & 7-70d & 2-10\,M$_\odot$ \\
Magnetar/NSAD & $1-3\times 10^4$y & 10-1000y & $\sim 10-30$\,pc & 0.01-1\,pc & 200-600y & 70-700d & 0.5-2\,M$_\odot$  \\
\hline
\end{tabular}
\caption{The final binary-induced mass loss phase is likely to occur between helium depletion (He Dep) and carbon ignition (C Ign).  Assume an ejecta velocity of $1000\,{\rm km \, s^{-1}}$, we infer the position of this ejecta shell.}
\label{tab:csm}
\end{center}
\end{table*}

If we assume an ejecta velocity of $\sim 1000 \, {\rm km \, s^{-1}}$ (roughly the escape velocity for helium stars), we can infer the rough position of this binary interaction ejecta and the timescale at which the hypernova shock would hit the shell.  If the binary mass transfer occurs at the end of helium burning, the shell will be sufficiently far from the star that it will not affect the light-curve.  But, especially for the BHAD engine, if the mass outflow occurs at the onset of carbon or oxygen burning, we expect strong shock interactions.  Some shock interactions should be expected as the hypernova ejecta propagates through the clumpy Wolf-Rayet winds~\citep{2020ApJ...898..123F}.

If the mass ejection occurs during helium depletion, 10,000 years before collapse, the shell of ejecta will integrate within the interstellar medium and it will be difficult to detect evidence for the mass loss.  However, if the mass ejection occurs after carbon ignition, the SN/jet shock will hit the shell within 7-70\,d.  The densities of these shells are on par with the explosive ejecta, $10^{-16}-10^{-12} \, {\rm g \, cm^{-3}}$, and the shock in the interaction will significantly decelerate the shock, converting kinetic energy into thermal energy.  In these scenarios, shock heating dominates the light curve.  It is possible that spectral features may allow us to distinguish different heating sources.  Simple models like those presented here predict very different properties of the light-curves and spectra but, as is often the case in nature, reality is much more complex.  Distinguishing these models may be very difficult.  One approach is to look at late-time observations to distinguish the energy sources~\citep{2021ApJ...918...89A,2022A&A...657A..64S,2024Natur.628..733R}.

\begin{figure}
    \includegraphics[width=0.5\textwidth]{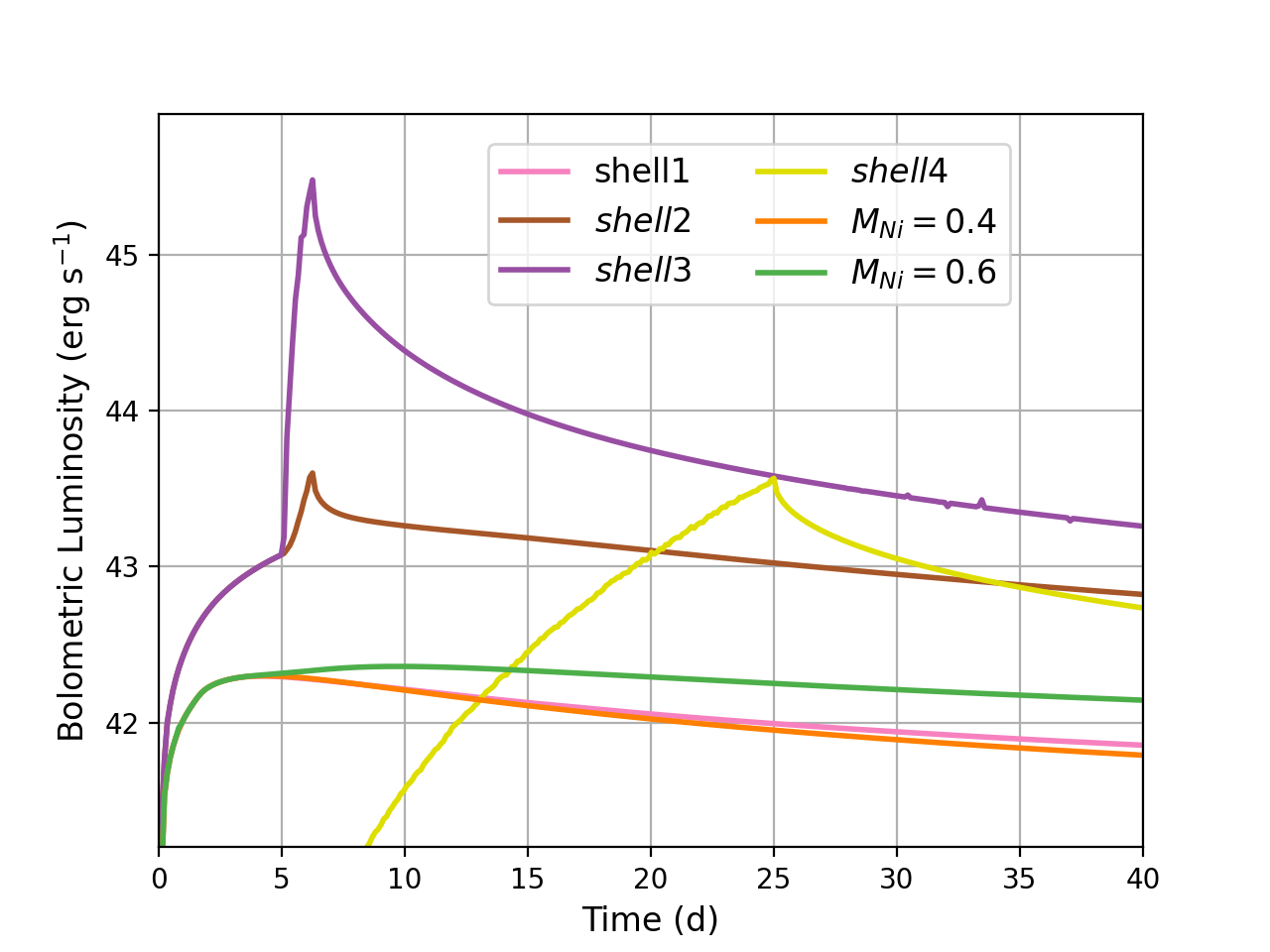}
    \caption{Comparing bolometetric light-curves from high nickel yield Ic-BL and low nickel yield Ic-BL with weak shocks.  We defer a more detailed study of the spectra and broad-band light-curves to an upcoming paper (Fryer et al. in preparation).} 
    \label{fig:grblc}
\end{figure}

Figure~\ref{fig:grblc} shows the SN light-curves for a select set of ejecta properties and nickel masses and mass distributions.  For the inner shells, it will be difficult to distinguish the peak emission from light-curves powered by $^{56}$Ni decay.  If the shocks are strong, shock interactions can make extremely bright SNe.  For shells that are sufficiently far out that the shock interaction with the shell occurs 20-30\,d after the launch of the explosion, we expect a double-peaked light-curve.  Observations or lack-thereof of this second peak will place constraints on our CO binary model.  Although there are some Ic-BL that show evidence of shock interactions~\citep{2014ApJ...782...42C,2019MNRAS.489.1110W}, the simple models used to infer the properties of the mass ejection (using assuming a wind-like mass loss) can produce inaccurate conclusions.  The structure of this ejecta, particularly the inhomogeneities, can alter the emission from these shock interactions~\citep{2020ApJ...898..123F}.  A broad wavelength coverage (from radio to X-ray) of the emission will likely constrain the structure, but a better understanding of shock interactions to determine the extent of these constraints. 

\subsection{Gravitational Waves}

The collapse of mildly rotating stars does not lead to strong gravitational wave emission.  The gravitational wave signal produced by the standard, convective-engine paradigm behind CCSN is a strong probe of the convection and its growth, but the gravitational wave amplitudes are typically 
$< 10^{-22} Hz^{1/2}$ at 10\,kpc~\citep[for a review, see][]{2011LRR....14....1F}.  Even for next-generation detectors like Cosmic Explorer~\citep{2019BAAS...51g..35R}, a 5-10~$\sigma$ observation is limited to a distance of $\sim 1$Mpc, our local group.  

Modestly rotating models, e.g. our rotating Kepler models and wide separation CO- or He-core binaries lead to an aspherical collapse that will produce signals that are 5-10 times higher amplitude than slowly-rotating models.  A small fraction of these rotating models may be detectable out to the Virgo cluster by Cosmic Explorer.

But, as we discussed with our magnetar engine models, our tightest CO-star binaries will have such fast spins that they will undergo dynamical instabilities.  For bar-mode instabilities, the requirement on the spin is:  $\beta = E_{\rm rot}/|W| > 0.27$.  Such high values for $\beta$ are achievable in our binaries with orbital separations near the Roche limit.  These systems would be detectable by Cosmic explorer beyond 100\,Mpc~\citep{2023ApJ...956...19F}.  

Although a lack of a gravitational wave detection of a Virgo-cluster SN would not preclude a tight CO-star binary progenitor, a detection would prove that high spin systems like those made by our tight binaries are produced.

\subsection{Dependence on Metallicity and, hence, Redshift}
\label{sec:metallicity}

For our different engine scenarios, the fraction of massive stars that form GRBs is affected differently by the metallicity, arguing for a different evolution with redshift.  The rate depends both on the shape of the initial mass function and the wind mass loss~\citep{2022ApJ...929..111F}.  There is growing evidence that the initial mass function begins to flatten as the metallicity drops below $0.01-0.1\,Z_\odot$~\citep{2021MNRAS.508.4175C}.  As it flattens, the fraction of BH forming systems increases.  In addition, mass loss from winds decreases with decreasing metallicity.  Weaker winds also allow more BH forming systems to form.  For our binary progenitors we expect the number of BHAD explosions to increase with decreasing metallicity and, hence, increase with redshift.  

In contrast,  NS engines (NSAD, magnetar) are produced by lower-mass stars (less than $\sim 20M_\odot$) which depend much less on mass loss.  Because the relative fraction of these stars decreases as the initial mass function flattens, under the tidally-locked progenitor mechanism, we'd expect the rate of explosions from these engines to decrease with decreasing metallicity and hence with increasing redshift.  

But many questions remain unanswered on the metallicity dependence of GRBs.  For example, the apparent lack of metallicity evolution in GRB hosts~\citep{2023ApJ...954...13G} points toward missing aspects in our model, likely due to our lack of understanding of stellar and binary evolution.

\subsection{Failed versus Strong Jets}

An important open question in our understanding of the lGRB-SN connection is whether all Ic-BL SNe are produced by jet-driven engines where the difference between lGRBs and Ic-BL SNe without GRBs is determined solely by the Lorentz factor of the jet either because the jet was weak when launched or the jet was unable to clean out a funnel because of its short duration and baryon loading slows the jet~\citep{2001sgrb.conf..144N,2007AIPC..924..108N,2008AJ....135.1136M,2012ApJ...750...68L,Modjaz_2020}.  In our tight-binary model, depending on the binary separation, we produce different sized disk (which varies both the jet power and duration).  The binary separation coupled with the mass of the collapsing star can explain the transition between lGRBs, LLGRBs and Ic-BL SNe without GRBs based on the correlation between .  The most powerful jets produce lGRBs.  These lGRBs would be produced by our tightest binaries and more massive BH-forming stars.  This high energy may also affect the associated SNe.   The current set of observed Ic-BL SNe accompanying lGRBs are very luminous. LLGRBs would be produced by wider binaries, preferentially from the low-mass BH-forming stars.  Ic-BL SNe without GRBs would be produced by our least powerful jets from the widest binaries and preferentially low-mass progenitors.  

Although we just described some basic trends with mass, without a more detailed understanding of the mass-loss mechanism forming CO binaries, it is difficult to make strong claims on these masses or rates.  Our poor understanding of the jet-engine and its propagation through the star also prevents solid constraints on the properties of the Ic-BL SNe produced in all of these cases.  Here we differentiate three scenarios for tight-binary, BHAD engines:    
\begin{itemize}
    \item Massive stars in the higher-mass range (above $\sim 30 \, M_\odot$) in tight binaries produce strong jets with a range of durations.  If the disk winds dominates the SN energy, more massive stars will produce SNe with higher mass, but higher total energy.  In this scenario, more massive stars will produce more massive and higher-energy Ic-BL SNe.  Wider binaries will produce weaker explosions, producing a range of energies from the same progenitor mass.  
    \item Massive stars on the lower-mass range (roughly in the $20-30 \, M_\odot$ range) produce weak jets (either LLGRBs or Ic-BL SNe without GRBs).  If these explosions are equally efficient at tapping jet energy to drive the SNe, they will be weaker than GRB SNe.  But it is likely that these jets inject more of their energy into their host stars and hence these SNe could have a range of energies.
    \item Wide binaries will also produce weaker GRBs and Ic-BL SNe without GRBs.  The weakest explosions will be lower-mass progenitors.
\end{itemize}

The extent of the broad lines may be able to help distinguish between these different scenarios.  The violent mass loss required in the formation of these tight binaries suggests that shock heating could dominate the light curve.  If so, the luminosity of the SN depends more on the circumstellar medium than it does on the energy in the SN explosion.  In this case, broad-band wavelength coverage (radio, Ultraviolet, X-ray) could provide clues to the nature of the shock interactions.  Nonetheless, any trends in the SN and GRB properties can be used to provide clues into both the progenitor binary-induced mass-loss and jet formation.

\subsection{Jet-Engine Duration and Baryon Loading}

The initial jet is typically baryon-loaded as it pierces through the star~\citep{1999ApJ...524..262M}.  The jet engine requires some time (some factor roughly equal to the jet transit time) to clear out a funnel to achieve high Lorentz factors (prior to clearing out this funnel, the jet gains mass, a.k.a. baryon loaded).  Although the front of the jet can accelerate at breakout~\citep{1968CaJPh..46..476C,2001ApJ...551..946T}, if it is baryon loaded, it can not reach high Lorentz factors.

For our C/O cores, the transit time of the jet through the star is less than 1\,s, much shorter than the BHAD engine lifetimes.  For the BHAD engine, our tight CO binaries should be able to clear out the jet region, allowing the creation of relativistic jets.  For more extended binaries, predominantly He-cores, we can have shorter-lived and weaker jets that may fail to reach high Lorentz factors.

Many models assume that the BHAD GRB duration is set by the accretion timescale of the disk.  Under this engine, our most massive progenitors or systems in wider separations would produce shorter-lived GRBs.  To produce longer-duration GRBs, we'd be limited to tight binaries with lower-mass CO cores or would have to invoke models like~\cite{1999Sci...284..115V} to extend the disk duration.  However, the duration for some models producing the gamma-ray emission may not be constrained by the jet engine~\citep{2011ApJ...726...90Z}.  In such a scenario, the duration of the jet engine is not an important constraint.

The duration of the engine dictates how baryon-free the jet becomes (determining the uppermost Lorentz factors of the jet).  This jet-engine duration is also linked to the duration of the GRB.  For all prescriptions of the jet-driven engines, the jet energy is proportional to the accretion rate.  The power of the jet can decrease dramatically with decreasing accretion rate.   For our more-massive stars, the duration of the jets is limited to less than 100\,s, but lower mass BH-forming progenitors can form longer-duration engines.  But, under the BHAD engine, these engines will be weaker beyond 100\,s.  Observations indicating that the jet weakens at late times, $\propto t^{\sim -1}$, would support a lower-mass BH forming progenitor.

\section{GRB Constraints}
\label{sec:GRB}

It has already been noted that observations indicating that lGRBs occur preferentially in low-metallicity, star-forming galaxies~\citep{2006Natur.441..463F} already argues that the engine under our tight-binary progenitor must be the BHAD engine.  But we can use additional observations to determine the role of the tight-binary progenitor scenario in explaining a wide range of GRB subclasses, including ultra-long and low-luminosity GRBs.

Note that we place GRBs 061208 and 100316D into the low-luminosity GRB class despite meeting our criterion of ultra-long GRBs, for reasons explained below.

\subsection{Ultra-long GRBs}

Ultra-long GRBs have been discussed as possibly arising from a unique progenitor compared with the broader lGRB population \cite[e.g.][]{levan2013new}. While the definition of ultra-long is imprecise, we adopt the definition of an observed duration in excess of 1000~s.  Some papers suggest ultra-longs are merely the extreme tail of the long population \citep[e.g.][]{virgili2013grb}. Other papers invoke alternative models which include a minor body falling onto a NS \citep{campana2011unusual}, a helium star-NS merger \citep{thone2011unusual}, the collapse of a blue supergiant \citep{gendre2013ultra,nakauchi2013blue}, a magnetar engine \citep{gompertz2017magnetars}, tidal disruption events \citep{ioka2016ultra}, and other unique origins \citep{kann2019highly}. 


These non-standard scenarios are typically invoked because it is thought to be difficult to explain the extreme length of prompt duration with a Wolf-Rayet progenitor. This relies on two assumptions: i) that the emission duration is of similar order as the accretion time, and ii) that the accretion time for a collapsar is effectively the free-fall time. We maintain the former, which is a prediction of internal shocks and some other prompt GRB emission models \citep{zhang2018physics}. The model described here challenges the second consideration, in some cases.  

From Figure~\ref{fig:mdot}, we find that the lowest-mass BH-forming progenitors in tight binaries produce the longest-lived accretion disks (due to the high angular momentum where the disk accretion time far exceeds the free-fall time).  The accretion rates in these disk steadily decrease with time and we expect lower jet powers as the ultra-long GRBs evolve.  Lower-mass progenitors should produce weaker jets on average.  We also expect ultra-long bursts to be less sensitive to metallicity than normal GRBs.  Finally, the low mass CO stars in these progenitors should be observed in any associated supernova with the lower masses tending to produce shorter supernova transients (if all else, e.g. shock heating, were equal). 

We can compare these predictions to the current observational sample of ultra-longs, compiled from the literature and summarized in Table~\ref{tab:grb_table_updated}. While the definition of ultra-long is imprecise, we adopt the definition of an observed duration in excess of 1000~s. For duration comparison with our accretion timescales we compare with the rest-frame duration.

Since luminosity is key, this restricts our sample to events with measured redshift. We compare the observed peak luminosity of our ultra-long sample to the broad sample of bursts reported in \citet{burns2023grb} which were detected by the Fermi Gamma-ray Burst Monitor (GBM) \citep{meegan2009fermi}. Using a two-sample Anderson-Darling test and find that they are incompatible at $>$99.9\% significance. While this is an imperfect test because of differential selection functions for different gamma-ray burst monitors, only two ultra-longs have peak luminosities above the median measured $L_{\rm iso}$ of $\sim5\times10^{52}$~erg. We thus confirm that these events appear to arise from jets with lower jet power, as compared with the broader population.

The highest $L_{\rm iso}$ ultra-long GRB is GRB~220627A, with a rest-frame duration of 900~s and a peak luminosity at the top 30th percentile. However, the ultra-long duration arises from a period of very weak emission, with large spikes ending only 300~s after the start (in the rest frame), which is consistent with our low-jet power picture at late times. There are a few GRBs whose overall rest-frame duration exceeds 1000~s; however, the only GRBs with impulsive behavior beyond this duration are GRBs 111209A, 130925A, and 090417B, with peak luminosities at the 1st percentile of the GBM sample. Thus, the durations and luminosities of ultra-long GRBs are consistent with model expectations.

\begin{table*}[t]
\centering
\begin{tabular}{|l|c|c|c|c|c|l|}
\hline
 & & Observed & Rest-Frame & & & \\
Burst & Redshift & Duration & Duration & $E_{\rm iso}$ & $L_{\rm iso}$ & References \\
\hline
GRB 220627A & 3.084 & 3700 & 900 & 2.30E+54 & 1.30E+53 & \citet{2022GCN.32295....1F} \\
GRB 101225A & 0.847 & 1377 & 800 & 2.70E+52 & 9.50E+52 & \citet{levan2013new} \\
GRB 121027A & 1.773 & 5700 & 2000 & 1.10E+53 & 1.20E+52 & \citet{levan2013new,lien2016third} \\
GRB 091024A & 1.0924 & 1200 & 600 & 3.50E+53 & 1.00E+52 & \citet{gruber2011fermi,virgili2013grb} \\
GRB 141121A & 1.469 & $\sim$1500 & $\sim$620 & 8.00E+52 & 1.00E+51 & \citet{2014GCN.17108....1G, cucchiara2015happy} \\
GRB 170714A & 0.793 & $\sim$1030 & $\sim$570 & 3.50E+52 & 8.30E+50 & \citet{lien2016third,2017GCN.21347....1P} \\
GRB 111209A & 0.677 & 10000 & 6000 & 6.00E+53 & 8.10E+50 & \citet{2011GCN.12663....1G} \\
GRB 130925A & 0.347 & 4500 & 3300 & 1.50E+53 & 4.00E+50 & \citet{2013GCN.15260....1G} \\
GRB 090417B & 0.345 & 2130 & 1600 & 4.50E+51 & 1.50E+50 & \citet{holland2010grb} \\
\hline
\end{tabular}
\caption{Table of Ultra-long GRBs and associated properties. Energetics values are either taken from bolometric values in papers or calculated from available flux/fluence and spectra values for this work.}
\label{tab:grb_table_updated}
\end{table*}

Thus far there are not dedicated studies on comparing a sample of ultra-long GRB host galaxy metallicities against the broader population. \cite{levan2013new} study three ultra-long GRBs and note the host galaxies may differ from the wider sample, but do not flag metallicity as an obvious difference. \citet{schady2015super} find a super-solar metallicity host for GRB~130925A, which is the second-longest rest frame duration GRB observed so far. Thus, this test is interesting but does not provide conclusive evidence either way, and motivates future host galaxy characterization of ultra-long GRBs.

Similarly, there are very few ultra-long GRBs with associated supernova. GRB~111209A, the longest GRB ever observed, has the associated SN~2011kl which appears more luminous than typical GRB supernovae \citep{kann2019highly}. Precise statements will require prioritized follow-up of future ultra-long GRBs to characterize their supernova.

\subsection{Low Luminosity GRBs}
An additional GRB type which is discussed as a separate class but may be part of a continuum are low-luminosity GRBs. They appear to arise from collapsars, being followed by Ic-BL SNe, but the specific cause of the lower GRB luminosity is not known. These events may be caused by off-axis jets akin to the low luminosity prompt GRB~170817A \citep{abbott2017gravitational}, but this is certainly not always the case \citep[e.g.][]{margutti2013signature}. Alternatively, the jet may barely escape the progenitor star which could result in a typical GRB event with lower overall power. Some of these may even fail to escape the star and be fully choked within them, producing a prompt high energy signal powered by a quasi-spherical shock breakout rather than internal dissipation of the energy within a jet. Indeed some low luminosity GRBs appear similar with such an outcome \citep{nakar2012relativistic}, but not all. The lowest luminosity events include GRBs 061208 and 100316D which appear to have very smooth, comparatively soft prompt emission, being consistent with expectations of shock breakout \citep[e.g.][]{campana2006association,margutti2013signature}. 

In addition to being followed by Ic-BL SNe, low luminosity GRBs prefer host galaxy metallicities similar to lGRB SNe, being much lower than the metallicities found in typical Ic SNe \citep{modjaz2020host}. This is strongly suggestive of a common progenitor of cosmological lGRBs. The non off-axis scenarios are all consistent with a common formation channel where the ultimate jet is weak, compared to cosmological GRBs. 

\section{SN Constraints}
\label{sec:SN}

SN observations can further constrain the nature of the progenitor and of the engine (or engines) responsible for the explosions.  Metallicity dependence for both Ic-BL associated with GRBs and those without GRBs can provide key insight into the engines and progenitors.  \citet{Modjaz2020} measured the gas-phase metallicities at the sites of 28 Ic and 14 Ic-BL SNe discovered by the Palomar Transient Factory, an untargeted wide-field optical survey. They also re-analyzed metallicity measurements of 10 GRB-SNe in the literature in a consistent fashion, of which six (GRB\,980425, XRF\,020903, GRB\,031203, GRB\,060218, GRB\,100316D, GRB\,120422A) are conventionally considered to be ``low-luminosity'' \citep{Cano2017}. They found that the Ic-BL SNe had statistically comparable metallicities to the GRB-SNe, but systematically lower metallicities than the Ic SNe.  This observation is consistent with the predictions of the BHAD engine.   

As we have shown in Section~\ref{sec:nuc}, the amount of $^{56}$Ni produced in our disk models depends upon the disk properties.  In the tight binary progenitor, the angular momentum in the disk increases with the BH spin.  For our current engines, we'd expect strong jets when the angular momentum is higher.  Because the $^{56}$Ni ejecta mass increases with angular momentum, by studying the $^{56}$Ni ejecta GRB SNe and correlating them with the power of the GRB, we can begin to probe the role of the accretion disk on the central engine.  Among GRB-SNe, no correlation has been found between gamma-ray energy and nickel mass (\citealt{Blanchard2024,Srinivasaragavan2024}; although see \citealt{li2006correlation}), with the caveat that most GRB-SN nickel masses are derived using the peak luminosity of the optical light curve \citep[]][; see below]{Cano2017}. GRB\,221009A \citep{Williams2023,Burns2023} was a particularly extreme case: despite having $E_{\gamma,\mathrm{iso}} \sim 10^{55}\,$erg, 7 orders of magnitude greater than GRB\,980425 \citep{Cano2017}, the SN was not more luminous than the SN associated with GRB\,980425 (SN\,1998bw), and there is no indication that it differed significantly from typical GRB-SNe \citep{Levan2023,Shrestha2023,Srinivasaragavan2023,Blanchard2024}.  Is the lack of correlation due to a stochasticity in the gamma-ray emission from jets, a lack of understanding of the jet mechanism or disk nucleosynthesis, or an issue with our $^{56}$Ni measurements?  Observations that better constrain the mass of $^{56}$Ni (e.g. late-time observations, particularly those extended to the IR which may contribute significantly to the bolometric luminosity; \citealt{lyman2014bolometric}) will allow us to understand the formation scenario behind tight binaries.  Improved disk models to better understand disk nucleosynthesis are also important in using this observational constraint.

One of the biggest uncertainties in our tight binary progenitor model is that we currently don't understand the method by which we eject the helium envelope.  Circumstellar interactions allow us to probe the nature of the binary interactions that will shed this mass.  If we see evidence of early interactions, the He-shell mass loss must have occurred after carbon ignition in the core.  If we can prove that the circumstellar interactions are minor in the observed light curve, the mass loss likely occurred just after helium depletion.  

In Table~\ref{tab:bts-icbl} we summarize current constraints on dense circumstellar interaction in Ic-BL SNe classified as part of ZTF's flux-limited experiment (the Bright Transient Survey; \citealt{Fremling2020,Perley2020}). We selected the subset of events with early $\sim$day-cadence data (specifically, a non-detection followed by a detection the next night, and another detection 1--2 nights later): 19 events of the total of 58\footnote{We do not include SN\,2019odp, which was subsequently reclassified as Type~Ib; \citealt{Schweyer2023}}. One of those events shows a very clear early blue peak that could be powered by interaction: SN\,2020bvc \citep{Ho2020,Izzo2020,Rho2021}. Another shows an entire light curve likely dominated by interaction: SN\,2018gep \citep{Ho2019,Pritchard2021}. This fraction (2/19) should be regarded as a lower limit, because other events may show more subtle signatures (e.g., not a full peak that rises then declines, but an ``excess'')---we will present the results of a detailed search in future work (Vail et al. in prep). We also perform forced photometry on ZTF images \citep{Masci2023} to search for late-time interaction signatures in the optical light curves, focusing on the objects at $z<0.04$ (where the ZTF limiting magnitude of 20.5\,mag corresponds to $M\approx-16\,$mag.

Detailed observations, including broadband coverage including radio and X-ray\citep[e.g.][]{2014ApJ...782...42C,Stroh_2021,2023ApJ...953..179C} of these interaction events will ultimately guide us to understanding better how the broad-line Ic progenitors are formed. 

\begin{table}[!h]
    \centering
    \begin{tabular}{l|c}
    \hline\hline
    Description & \# \\
    \hline
    BTS Total & 58 \\
    ... Subset with early $\sim$day-cadence data & 19 \\
    ... ... Early fast peak (double peaked) & 1 \\
    ... ... Light curve peak dominated by interaction & 1 \\
    ... Subset at $z\leq0.04$ & 34 \\
    ... ... Prominent late-time interaction peak & \textbf{0---temp.} \\ 
    \end{tabular}
    \caption{Signatures of interaction in broad-lined Ic supernovae from ZTF's Bright Transient Survey. The number of objects with clear interaction signatures in the optical light curve should be viewed as lower limits; searches for more subtle signatures will be presented in future work. \textbf{Jada checking the late-time light curves}}
    \label{tab:bts-icbl}
\end{table}

\section{Conclusions}

In this paper, we have studied the properties of the tight-binary progenitor for GRBs under three potential engines:  BHAD, NSAD, and magnetar.  We compare the predicted observational features of these progenitors and engines with the observed Ic-BL SN and GRB properties.  The primary results of these comparisons are:
\begin{itemize}
    \item The preference of lGRB engines toward lower metallicity argues that the BHAD engine is the primary GRB engine under the tight binary progenitor scenario.
    \item The fact that the broader Ic-BL SNe category also exhibits this same trend suggests that the same BHAD engine drives these explosions.
    \item lGRB, ultra-long GRB, LLGRB and Ic-BL can all be explained under the tight-binary progenitor where weaker and lower Lorentz-factor outbursts are caused by binaries with the lower-mass BH formation progenitors or progenitors with wider separations.
    \item Observations of shock interactions can guide our understanding of the mechanism behind mass-loss that creates these tight binaries
    \item Nucleosynthetic yields from these events can provide insight into the disk properties and, in particular, the behavior of disk winds in this BHAD engine
\end{itemize}

However, a number of additional progenitor/engine scenarios exist that have the potential to explain these observations.  For example, the Induced Gravitational Collapse and its related Binary-driven Hypernova model~\citep{2012ApJ...758L...7R,2014A&A...565L..10R} which invoke the collapse of a neutron star to a black hole also argued for a tight binary between a neutron star and a CO star.  In this model, when the CO star collapses and then explodes, the accretion of its ejecta onto the neutron star causes it to collapse.  To get sufficient accretion to produce a black hole, the binary must be extremely tight which can only occur if the collapsing star is a CO, not a Helium star~\citep{2014ApJ...793L..36F,2015ApJ...812..100B,2024arXiv240115702B}.  Multi-dimensional models of these exploding binaries exist~\citep{2015ApJ...812..100B,2024arXiv240115702B}, but they have not been followed through light-curve calculations.  As such, it is not clear whether these explosions will match the observed Ic-BL SNe.

Another progenitor is the extended mixing model that produces homogeneous stars tends to occur preferentially in high mass stars~\citep{2006NCimB.121.1631Y,2013ApJS..204...16F} and produce more BHAD systems with lower metallicity~\citep{2006NCimB.121.1631Y}.  This progenitor paradigm also suggests that BHADs are the dominant engine.  It is likely that, if strong mixing occurs, the associated supernova-like transient produced in this progneitor will tend to be Ic supernovae (with very little He and H in the ejecta).  It is less clear what this progenitor predicts for the circumstellar medium or the final rotation speeds which still dependent on the strength of coupling terms like the Spruit-Taylor dynamo.  

The Helium merger model~\citep{1998ApJ...502L...9F,2001ApJ...550..357Z} will produce extremely rapidly-spinning disks and could even explain the ultra-long GRBs~\citep{2011Natur.480...72T}.  It will certainly have nearby mass-ejection shells in the circumstellar medium that will affect the supernova light-curves.  But it is less clear how this model will not produce more type Ib over type Ic supernovae.  Both the helium-merger and homogeneous star progenitors must be studied in more detail to test against the observations.

Further observations are critical to improving our understanding of the engines and progenitors of lGRBs, ultra-long GRBs, LLGRBs, and Ic-BL SNe.  These include:
\begin{itemize}
    \item Improved observations of GRB durations, jet energies and Lorentz factors as a function of environment properties (metallicity, host-galaxy characteristics) and GRB sub-type.  The different GRB subtypes depend on the progenitor mass and binary separation.  Observations that accurately constrain the jet properties can better uncover the nature of these different explosive phenomena.
    \item Trends in Ic-BL energetics  comparing both GRB-SNe and normal Ic-BL to normal type Ic SNe.  Trends in the energetics will provide insight into the jet-mechanism and how it interacts with the collapsing star.
    \item Trends in the Ic-BL nucleosynthetic yields (e.g. $^{56}$Ni) between different Ic classes.  Peak luminosities may are not be ideal constraints on the yields and late-time or other observational methods must be developed to understand these yields.  This includes helium that is either in the explosion or swept up in the explosion.
    \item Evidence or lack thereof of shock interactions.  This probes the timescale of mass loss, providing insight into the mechanism behind binary-interaction He mass-loss.  This includes broad-band observations including radio and X-ray.
    \item High energy monitors built to target ultra-long and low-luminosity GRBs are key to building the observational database to test our unified picture.
\end{itemize}
These observations must be coupled with advanced theoretical understanding including:
\begin{itemize}
    \item Better understanding of binary interactions to include He-star mass loss in population studies of binaries
    \item Better understanding of disk wind and jet properties
    \item Improved shock interaction studies to better take advantage of improved observations of these interactions.
\end{itemize}
With advances in observations and theory, we can both confirm/refute the tight-binary progenitor model and better understand the physics behind these powerful explosions.

\begin{acknowledgements}

The work by CLF was supported by the US Department of Energy through the Los Alamos National Laboratory. Los Alamos National Laboratory is operated by Triad National Security, LLC, for the National Nuclear Security Administration of U.S.\ Department of Energy (Contract No.\ 89233218CNA000001).  This work was performed in part at Aspen Center for Physics, which is supported by National Science Foundation grant PHY-2210452. A.C. acknowledges support from the National Science Foundation via the grant AST-2431072, and from NASA via several \textit{Swift}/GI awards.

\end{acknowledgements}

\bibliography{refs}{}
\bibliographystyle{aasjournal}

\end{document}